\documentclass[12pt]{article}
\usepackage{graphicx} 
\usepackage{amsmath}
\usepackage{lmodern}
\usepackage[utf8]{inputenc}
\usepackage{cite}
\usepackage{color}
\title{ Entropy, purity and gluon cascades at high energies with recombinations and transitions to vacuum }

\author{Krzysztof Kutak$^{1,2}$, Michał Praszałowicz$^{3}$}
\date{
  {\small
  $^{1}$ \it{Institute of Nuclear Physics, Polish Academy of Sciences, ul.~Radzikowskiego 152, 31-342, Krak\'ow, Poland}\\%
  $^{2}$ \it{CPHT, CNRS, Ecole Polytechnique, Institut Polytechnique de Paris, 91120 Palaiseau, France}\\
  $^{3}$ \it{Institute of Theoretical Physics, Jagiellonian University, Lojasiewicza 11, 30-348 Kraków, Poland}
  }
}

\begin{document}

\maketitle
\begin{abstract}
We study one dimensional 
dipole cascade models in the high-energy limit of QCD. Motivated by data on hadron multiplicities in the LHCb kinematical range, we generalize existing cascade models for splitting and recombination to account also for transitions to the vacuum. This modification allows us to describe the data. Furthermore, we perform both analytical and numerical studies of the cascades and find that the cascade including loop corrections, as well as the one accounting for transitions to the vacuum, can both be related to the Negative Binomial Distribution. In the latter case, however, one must properly normalize the cascade to account for transitions to the vacuum. We also study the scaling properties of the cascades and identify new regimes, which we call "focal" and "parallel." Finally, we investigate the Quantum Information (QI) measures of the cascades, which allow us to highlight their distinctive properties.
\end{abstract}

\medskip

\section{Introduction}
With recent advances in quantum information and quantum computing, high-energy physics has begun to ask questions similar 
to those asked in the foundations of quantum mechanics \cite{Afik:2025ejh}.  These questions concern manifestations of entanglement 
that can be observed in various correlations of measured quantities. In high-energy physics, 
such correlations are studied in processes involving the top quark, Higgs boson decay, $\alpha$ polarization,
and multiparticle production, spin correlations in quark-antiquark production and relation of spin and angular momentum \cite{Afik:2025ejh,DelGratta:2025qyp,STAR:2025njp,Maltoni:2024csn,Aoude:2023hxv,Datta:2024hpn,Florio:2025hoc,Qi:2025onf,Hatta:2024lbw,Altomonte:2024upf}. 
Recently, it has been proposed that entanglement can be studied in the process of Deep Inelastic Scattering (DIS)
 where electron probes a proton with a virtual photon \cite{Kharzeev:2017qzs}.
In isolation, a proton is a coherent, pure quantum state. Interaction destroys quantum coherence, and moreover, in 
DIS certain degrees of freedom are effectively integrated out,  leading to the formation of a mixed state. 

This process has been studied in the high-energy limit of QCD \cite{Kharzeev:2017qzs} where it is natural to formulate it using
Mueller dipole picture \cite{Mueller:1993rr,Mueller:1996te}. The proton wave function is constructed from color dipoles and as one goes 
to higher and higher energy, more and more dipoles are produced. By tracing out the quarks, keeping the size of the initial dipole fixed, and integrating over the color and transverse sizes of the daughter dipoles, the large-$x$ degrees of freedom become entangled with the low-$x$ degrees of 
freedom — represented by gluons (or by dipoles in Mueller's approach).
This procedure leads to the emergence of a so-called reduced density matrix
characterizing soft gluons {\em i.e.} low $x$ degrees of freedom \cite{Liu:2022hto} and the corresponding partonic entanglement entropy. 
The entries of the density matrices are given by dipole multiplicities which obey the low $x$ evolution equations \cite{LevinLublinsky:2003Linear}.

In Ref.~\cite{Kharzeev:2017qzs} it has been conjectured that the partonic entanglement entropy is directly related 
to the entropy of produced hadrons. This picture has been largely confirmed by recent studies of the DIS data, both fully inclusive and diffractive \cite{Hentschinski:2022evidence,Hentschinski:2022maxentDIS,Hentschinski:2024qcd_evo,Hentschinski:2023maxent}, 
and there are indications that it holds for the $pp$ case as well \cite{Tu:2019ouv,Datta:2024hpn}. 
In the high-energy limit entropy has been found to grow linearly with rapidity \cite{Kharzeev:2017qzs}, which confirms earlier results \cite{Kutak:2011rb}
(see also Ref.~\cite{Kutak:2023cwg} for a review) indicating maximal entanglement and signaling conformal properties of QCD at high energies \cite{Gursoy:2023hge}. 
For other papers addressing the entanglement entropy in high-energy collisions see 
\cite{Peschanski:2012cw,Stoffers:2012mn,Gotsman:2020bjc,Kovner:2015hga,
Kovner:2018rbf,Peschanski:2016hgk,Peschanski:2019yah,Dumitru:2023fih,Dumitru:2023qee,
Ehlers:2022oal,Dumitru:2025bib,Berges:2017hne,Dvali:2021ooc,Chachamis:2023omp,Rybczynski:2025ccy,Ramos:2025tge,Cheskis:2025mrs,Ouchen:2025tta}.

The proton-proton case is more complex than DIS, because both the projectile and the probe are bound states characterized by strong interactions, Furthermore, the bulk of final states is produced in the central rapidity region where one can not  easily define projectile and target. 
However, one may consider such measurements in which the partonic systems of protons are probed in a phase space region  where one proton
may be considered as a dilute system of partons, while the other as a dense one. Such processes may be addressed within the framework
of Mueller's one dimensional dipole cascades \cite{Mueller:1996te}, which are believed to mimic the four dimensional QCD evolution.
Different forms of such cascades have been studied in 
Refs.~\cite{Kharzeev:2017qzs,Gotsman:2020bjc,Iancu:2004iy,Iancu:2005dx,Shoshi:2005pf,Kovner:2005aq,
Kozlov:2006zj,Bondarenko:2006rh,Kovner:2024tin,Braun:2024dfr,Kou:2022dkw}.

Throughout this paper we  use the simplest version of the reaction-diffusion process \cite{Iancu:2004iy}, which was shown to follow from the Reggeon field
theory with 3-Pomeron and 4-Pomeron couplings \cite{Bondarenko:2006rh}.  The resulting cascade can be characterized by scaling 
that emerges from an interplay of the production part and the recombination part \cite{Iancu:2004iy}. 
We carry numerical studies of the splitting-recombination cascades  and find a new regime of this scaling. We also show that
the resulting probability distribution is to a very good approximation a Negative Binomial Distribution (NBD) commonly used
to describe multiplicity distributions in hadron production \cite{Giovannini:1985mz}. We show that the cascade multiplicity 
and the NBD parameter $k$ first rise with increasing energy and then saturate at a fixed, relatively large value. 
However, the probability distribution never reaches the Poisson limit.
Rising $k$ is in flat contradiction
with the LHC multiplicity data \cite{Praszalowicz:2011zza} where $k$ decreases with energy. Inreasing
$k$ naturally follows from many multiparticle production models 
such as glittering glasma \cite{Gelis:2009wh} or string percolation model \cite{DiasdeDeus:2010ggs}.

To circumvent this discrepancy, we generalize the parton cascade model to include not only dipole recombination but also dipole transition to vacuum.
Such transitions appear in the RFT \cite{Bondarenko:2006rh}, although the probabilistic interpretation is hampered by the negative sign 
of merging probability~\cite{Bondarenko:2006rh,Braun:2024dfr}. In the generalization  proposed here all probabilities are positive.
Unfortunately, the behavior of the NBD $k$ parameter is qualitatively not changed, however, the multiplicity does no longer saturate and tends to zero for large
$y$. In this way the system is pulled into a zero particle state, but this limit is achieved with constant $k$.

We further discuss quantum measures that allow us to characterize properties of the dipole cascades. 
Those are Krylov complexity, variance, and most importantly entropy and  purity.  The study follows earlier work \cite{Caputa:2024xkp}.

We attempt to use the splitting-recombination cascade to describe hadronic 
entropy as can be obtained from the LHCb measurement of forward hadron production. 
However, the recombination is insufficient. One has to add an 
additional term (discussed above) which effectively models transition of dipoles to states that are not measured or to vacuum. 
Whith such modification we achieve  description of entropy data. Furthermore, we find its approximate analytical 
solution which again can be related to the NBD  but with a new term corresponding to probability of particles recombining to vacuum.

The paper is organized as follows: in Sect.~\ref{sec:equations} we formulate the cascade equations and relate
the resulting probabilities to
geometric and negative binomial distributions. Then in Sect.~\ref{sec:numcascades} we study numerical results
for the cascades introduced in Sect.~\ref{sec:equations}, identify different evolution regimes, analyze the
the probability distributions and study the resulting scaling properties. Sect.~\ref{sec:EQI} is devoted to
the quantum measures of the multiplicity distributions following from the cascade equations, 
and in Sect.~\ref{sec:LHCb} we apply the formalism
developed in the previous sections to the LHCb experimental data on particle multiplicities in the forward 
rapidity region. Discussion and conclusions are presented in Sect.~\ref{sec:conclusions}.

\section{Dipole evolution equations}
\label{sec:equations}

 A convenient framework
to describe multiple particle production in the low $x$
limit is provided by the color dipole picture \cite{Mueller:1993rr,Mueller:1996te}, for
a review see \cite{Kovchegov:2012mbw}. In this approach, which was originally formulated for DIS, 
the initial dipole is created as a result of the splitting of a virtual photon into a 
quark-antiquark pair. Then, as the energy increases, more and more dipoles are produced.
This cascade  can be then used to  describe the multiple-particle production. It corresponds 
to the initial state radiation or, in the Monte Carlo language, to the initial state shower. 
Starting from the QCD in the large number of colors ($N_c$) approximation 
one can derive equation for dipole probability distribution $p_n(y, {r})$, which takes the following generic
form \cite{Mueller:1993rr},

\begin{align}
\frac{\partial p_n(y, \{r\})}{\partial y} = \sum_{m} K \otimes p_m(y, \{r\})
\end{align}
In this equation, $K$ is the kernel of the Balitsky, Fadin, Kuraev, Lipatov (BFKL) \cite{Balitsky:1978ic,Kuraev:1977fs} evolution written in two-dimensional position space, 
and $r$ refers to the size of dipoles in transverse space.
This equation can then be used to obtain the BFKL equation in the linear regime and the Balitsky, Kovchegov (BK) equation \cite{Balitsky:1995ub,Kovchegov:1999yj}
if multiple scattering and/or recombination are taken into account.

Recently,  one dimensional reduction of the equation for dipole multiplicities \cite{LevinLublinsky:2003Linear} become attractive 
due to its simplicity and phenomenological success in describing entropy of produced hadrons (see {\em e.g.} \cite{Kharzeev:2017qzs})
\begin{equation}
\partial _{y}p_{n}(y) =-\alpha np_{n}(y)+\alpha (n-1)p_{n-1}(y)  \, .
\label{eq:Equation0}
\end{equation}
Here, $\alpha$ corresponds to the splitting kernel which in the one-dimensional case  is just a number. 
It directly corresponds to the Pomeron intercept. In the BFKL case 
$\alpha=4 \ln 2 (N_c\alpha_{\rm s}/\pi)$, where $\alpha_{\rm s}$ is a strong coupling constant, here, it has to be 
either fitted or taken as the BFKL value.

The interpretation of the terms entering Eq.\thinspace(\ref{eq:Equation0}) is
rather straightforward. The probability of finding $n$ partons (dipoles) in the system is $p_{n}$ and
the total probability that the system transforms into a configuration of $n+1
$ dipoles is $\alpha\,n\,p_{n}$. Such a process reduces the probability to
find $n$ partons, therefore it enters with a minus sign. Additionally,
the system of $n-1$ dipoles can transform into $n$ dipoles by splitting $%
1\rightarrow 2$, increasing the probability to find $n$ partons, and the
probability is in this case $\alpha (n-1)p_{n-1}$. It enters with the plus
sign because it corresponds to the increase in the number of partons (gain).
One can explicitly check that the normalization $\sum_n p_n=1$ is preserved in the course of evolution.

The analytical solution to Eq.~(\ref{eq:Equation0}) is known already
for some time (see {\em e.g.} \cite{Mueller1995UnitarityBFKL}), as it corresponds to the BFKL evolution where
the average multiplicity $\bar{n}$ grows exponentially with $y$%
\begin{equation}
\bar{n}(y)=e^{\alpha y},
\label{eq:nbarBFKL}
\end{equation}
which corresponds to the powerlike growth with energy. 
The probability distribution is
also known and is given by the geometric distribution%
\begin{equation}
p_{n}(y)=\frac{1}{\bar{n}}\left( \frac{\bar{n}-1}{\bar{n}}\right) ^{n-1} \, ,
\label{eq:pdistr}
\end{equation}%
where $n=1,2,\ldots$. 
Note that using the standard notation for the geometric distribution 
\begin{equation}
P_{m}^{\text{G}}(\bar{n}_{\rm G})=\frac{1}{\bar{n}_{\rm G}+1}\left( \frac{\bar{n}_{\rm G}}{\bar{n}_{\rm G}+1}%
\right) ^{m} \, , \label{eq:trueg}
\end{equation}%
where $m=0,1,2,\ldots $ and $\bar{n}_{\rm G}$ is the average multiplicity, we have
\begin{equation}
p_{n}(\bar{n})=P_{n-1}^{\text{G}}(\bar{n}_{\rm G}=\bar{n}-1).  \label{eq:pneqPnG}
\end{equation}%
where $n=1,2,\ldots $. Evolution (\ref{eq:Equation0}) starts from $p_1(0)=1$,
$p_{n>1}(0)=0$
and never produces a nonzero $p_0$, while the geometric distribution
(\ref{eq:trueg}) includes $P_0^{\rm G}$. One can easily check that (\ref{eq:pneqPnG}) is properly normalized\footnote{See also study of 1 D model in  \cite{Liu:2022hto} where the lowest multiplicity is in fact $p_0$}.

The geometric distribution corresponds to a special limit of the negative binomial
distribution (NBD)%
\begin{equation}
P_{m}^{\text{NBD}}(k,\bar{n}_{\rm NBD})=
\frac{\Gamma(m+k)}{m!\,\Gamma(k)}
\left( \frac{\bar{n}_{\rm NBD}}{k+\bar{n}_{\rm NBD}}\right) ^{m}
\left( \frac{k}{k+\bar{n}_{\rm NBD}}\right) ^{k} \, ,
\label{eq:NBD}
\end{equation}%
where $m=0,1,2,\ldots$.
Indeed, for $k=1$ we have%
\begin{equation}
P_{m}^{\text{G}}(\bar{n}_{\rm G})=P_{m}^{\text{NBD}}(1,\bar{n}_{\rm G}).
\label{eq:PGvsPNBD}
\end{equation}
For $k\rightarrow \infty$ 
\begin{equation}
    P_{m}^{\text{NBD}}(k\rightarrow \infty,\bar{n}_{\rm NBD})=P_m^{\rm Poisson}(\bar{n}_{\rm NBD}) \, .
\end{equation}

The negative binomial distribution is fully determined by two parameters $\bar{n}_{\rm NBD}$
and $k$. The parameter $k$ is related to the variance \cite{Szwed:1987vj}
\begin{equation}
    \sigma^2_{\rm NBD}=\sum_{m=0}^{\infty}m^2 P_{m}^{\text{NBD}}(k,\bar{n}_{\rm NBD})-\bar{n}_{\rm NBD}^2
\end{equation}
in the following way
\begin{equation}
    k=\frac{\bar{n}_{\rm NBD}^2}{\sigma^2_{\rm NBD} -\bar{n}_{\rm NBD}} \, .
    \label{eq:kNBD}
\end{equation}
In other words, $P_{m}^{\text{NBD}}$ is determined by the first two moments.

As the energy increases, it becomes necessary to take into account the unitarity corrections. 
In general, this is a complicated problem, leading to nonlinear evolution equations that are nonlocal in transverse dimension \cite{Kovchegov:2012mbw}. 
In the 1D scenario, one method is to take into account the terms corresponding to dipole fusion. 
The resulting equation reads \cite{Iancu:2004iy}
\begin{align}
\partial_{y} p_{n}(y) &= -\alpha n p_{n}(y) + \alpha (n-1) p_{n-1}(y) \notag \\
                      &\quad + \beta n(n+1) p_{n+1}(y) - \beta n(n-1) p_{n}(y)
\label{eq:EquationSat1}
\end{align}
where the $\beta
\simeq \alpha \alpha _{\text{s}}^{2}$.
The equation models contribution corresponding to pomeron loop diagrams \cite{Bondarenko:2006rh,Braun:2024dfr} 
and slows down  the increase of mean multiplicity. 

The interpretation of $\beta$ terms entering Eq.~(\ref{eq:EquationSat1}) 
is straightforward.
The system of $n+1$ partons can transform by the annihilation
process $2\rightarrow 1$ into $n$ parton configuration. The probability to
find $n+1$ partons is $p_{n+1}$, and annihilation proceeds by picking
one parton out of $n+1$, and the second one out of remaining $n$ partons, which
gives an increase of probability of finding $n$ partons by $\beta
n(n+1)p_{n+1}$, hence a $+$ sign. In principle there should be a factor $%
1/2$ in front to avoid double counting, however, since we will treat $\beta $ as a free
parameter, it can be absorbed into $\beta $. Analogously, annihilation of two
partons out of $n$ reduces the probability, and the emission probability, in
this case $\beta n(n-1)p_{n}$, enters with a $-$ sign.
It is easy to show that also in this case $\sum_n p_n=1$.

This equation has been already studied in the literature
where it has been explicitly demonstrated using approximated analytical solution valid at large rapidity \cite{Hagiwara:2017uaz}, 
and numerical methods valid at any rapidity that the mean multiplicity saturates \cite{Caputa:2024xkp}. 
We extend the analysis of \cite{Caputa:2024xkp} and explicitly demonstrate this in the next section, where we will also argue that the probability distribution generated by~(\ref{eq:EquationSat1}) is very close to the NBD,
\begin{equation}
        p_n = P_{n-1}^{\rm NBD}(k,\, \bar{n}_{\rm NBD} = \bar{n} - 1),
\end{equation}
with $k$ given by~(\ref{eq:kNBD}). It will also turn out that $k$ saturates for large $y$.

Since experimentally, at least in $pp$ collisions at the LHC \cite{Praszalowicz:2011zza}, $k$ decreases with energy, 
we have generalized
Eq.~(\ref{eq:EquationSat1}) to the case where two dipoles can annihilate to vacuum, $2\rightarrow0$. Such
transitions appear in the Reggeon effective field theory \cite{Bondarenko:2006rh}. They can also mimic
disappearance of particles from the detector or as regions of phase space that are not available in the considered measurement.

To this end we add two more terms proportional to a new parameter
$\gamma$. By analogy with the
previous considerations, we have a loss term where $n$ partons transform
into $n-2$ configuration with probability 
$
-\gamma n(n-1)p_{n}
$,
and a gain term where $n+2$ partons transform into $n$ parton configuration with probability
$
+\gamma (n+1)(n+2)p_{n+2}
$.
The resulting equation reads as follows
\begin{eqnarray}
\partial _{y}p_{n}(y) &=&-\alpha np_{n}(y)+\alpha (n-1)p_{n-1}(y)  \notag
\\
&&+\beta n(n+1)p_{n+1}(y)-\beta n(n-1)p_{n}(y)  \notag \\
&&+\gamma (n+1)(n+2)p_{n+2}(y)-\gamma n(n-1)p_{n}(y)\, .
\label{eq:eqsat2}
\end{eqnarray}
One can check that this equation conserves probability. 
To some extent, similar considerations prompted the authors of the Ref.~\cite{Hentschinski:2024qcd_evo}  to generalize
 the 1 D model to account for measurements in the moving 
rapidity window. One necessarily had to introduce $p_0$ that was accounting for probability that particles are not 
detected in certain region of rapidity. The difference is that in the present paper such term is a part of 
evolution equation and does not have to be modeled independently.

\section{Properties of dipole cascades}
\label{sec:numcascades}

In this section, we perform a thorough numerical analysis of Eqs.~(\ref{eq:EquationSat1}) and (\ref{eq:eqsat2}).
It has been argued in Ref.~\cite{Blaizot:2006wp} that Eq.~(\ref{eq:EquationSat1}) has an approximate attractor
in a sense that the multiplicity moments of order $k$ scale as $(\alpha/\beta)^k$. We therefore define parameter
$r$
\begin{equation}
    \beta=r \, \alpha \, .
\end{equation}
In the case of 1 D Reggon  model of high energy limit of QCD we expect $r\sim \alpha_{\rm s}^2$. However, in the case of $pp$ scattering $\alpha_s$ can be quite large,
and $r$ may be of the order 0.05 or larger. In what follows we  vary $\alpha =0.2 \div 0.8$ and $r=0.05 \div 0.5$. 
For smaller $r$ we need better numerical accuracy. Our initial condition will be always $p_1(0)=1$ and $p_{n>1}=0$.
In each case we check whether the probability distribution generated by the differential equations 
(\ref{eq:Equation0}), (\ref{eq:EquationSat1}) and (\ref{eq:eqsat2}) are properly normalized.

\subsection{Cascade $\beta$}
\label{sec:beta}

\begin{figure}[h!]
  \centering
    \centering
 \includegraphics[width=7.5cm]{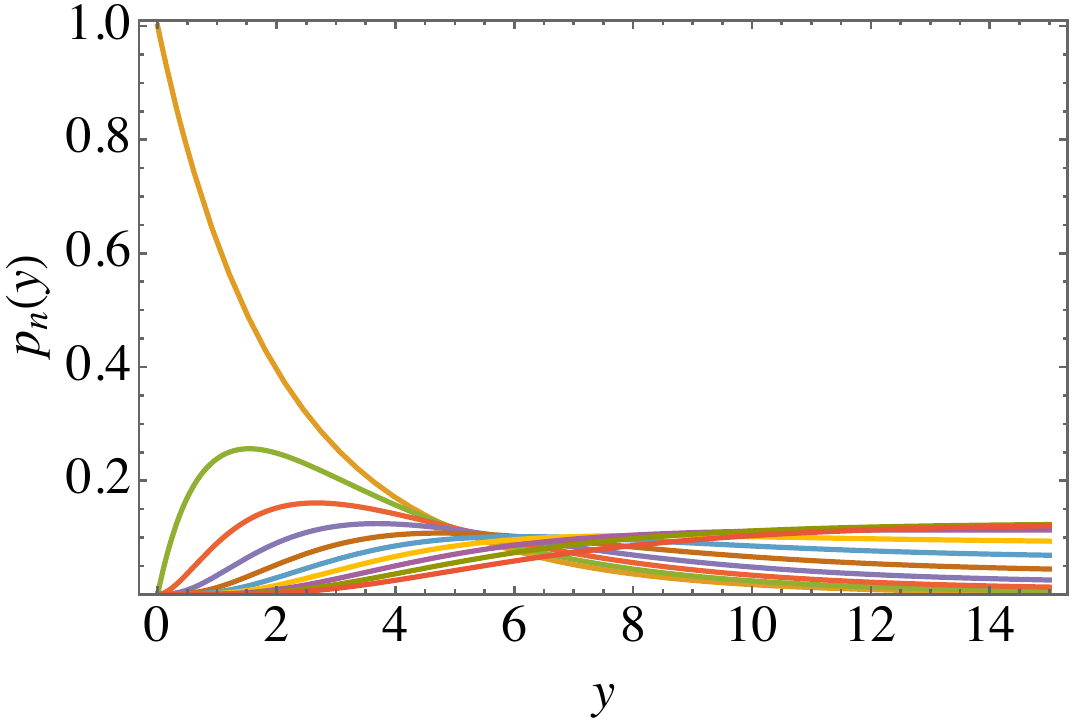}~
 \includegraphics[width=7.5cm]{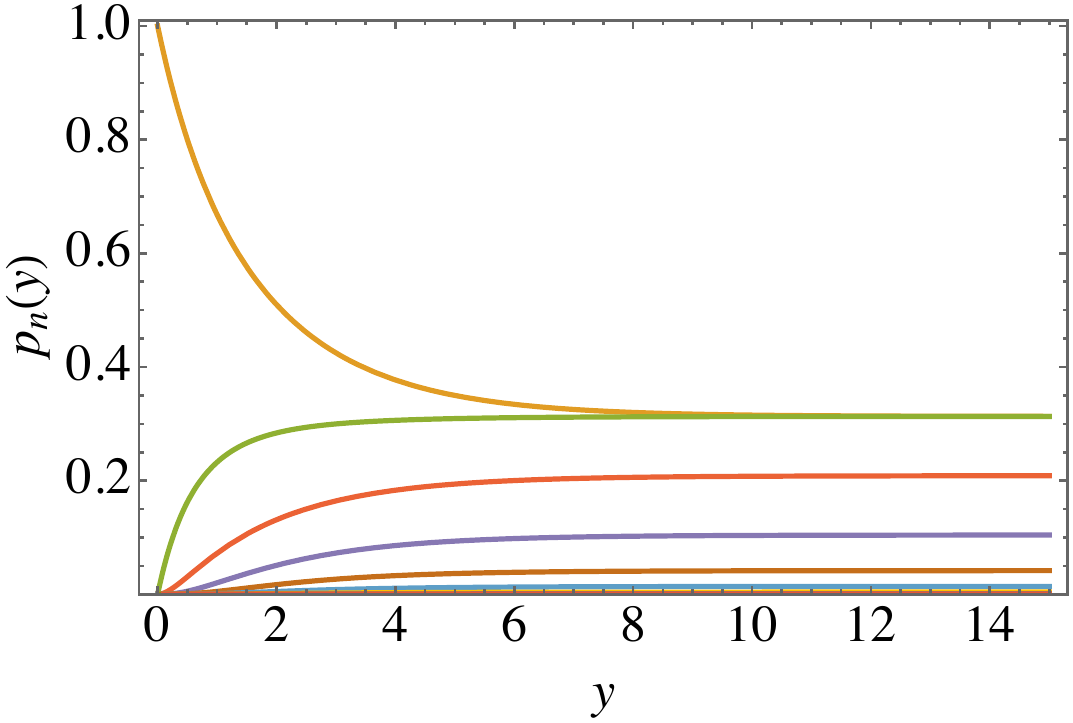}  
 \caption{Probabilities $p_n(y)$ ($n=1,\ldots ,10$) for the $\beta$ branching (\ref{eq:EquationSat1})
 for $r=0.1$ (left)  
 and $r=0.5$ (right), and $\alpha=0.5$.
 One can see that for small $r$ probabilities cross at $y_{\rm cross}\simeq 6.5$, whereas for larger $r$ they 
 reach asymptotic values without crossing.}
  \label{fig:pnyr}
\end{figure}

Let us first consider Eq.~(\ref{eq:EquationSat1}) which we will call the $\beta$-cascade, in contrast to the $\alpha$-cascade
of Eq.~(\ref{eq:Equation0}) and the $\gamma$-cascade of Eq.~(\ref{eq:eqsat2}). In Fig.~\ref{fig:pnyr} we plot first 10
probabilities for $\alpha=0.5$, and $r=0.1$ and $0.5$. We see that for small
$r$ probabilities cross in the  vicinity of some rapidity, which we call $y_{\rm cross}$, and then rearrange. On the other hand, for larger $r$'s
probabilities do not cross and are almost parallel for large values of $y$. 
In what follows, we will refer to these two regimes as {\em focal} and {\em parallel}, respectively.
The physical region of $\alpha$ and $\beta$ couplings relevant to QCD evolution is in the focal regime.

\begin{figure}[h!]
  \centering
    \centering
 \includegraphics[width=7.5cm]{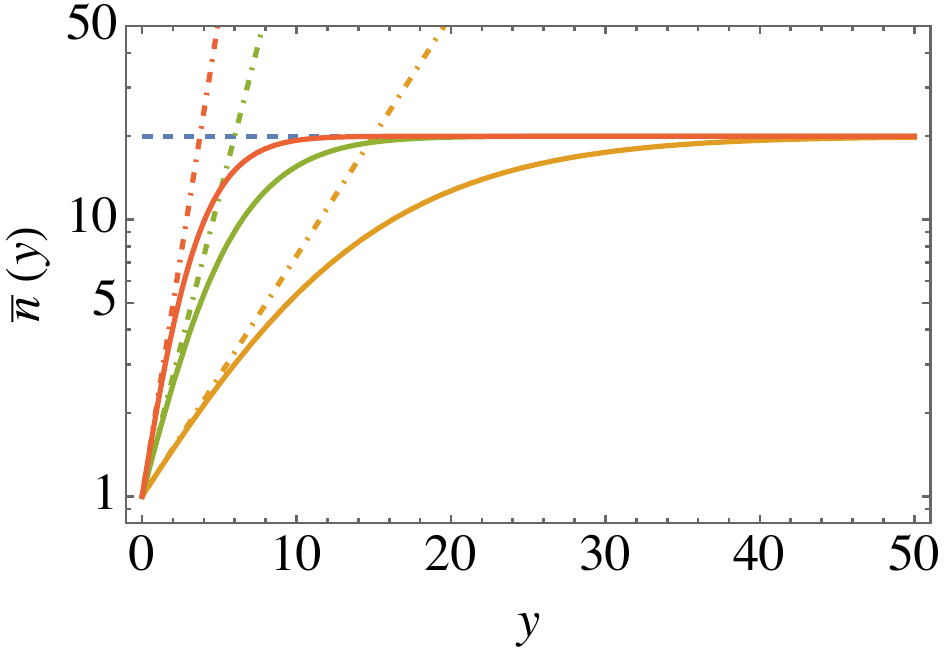} ~
 \includegraphics[width=7.5cm]{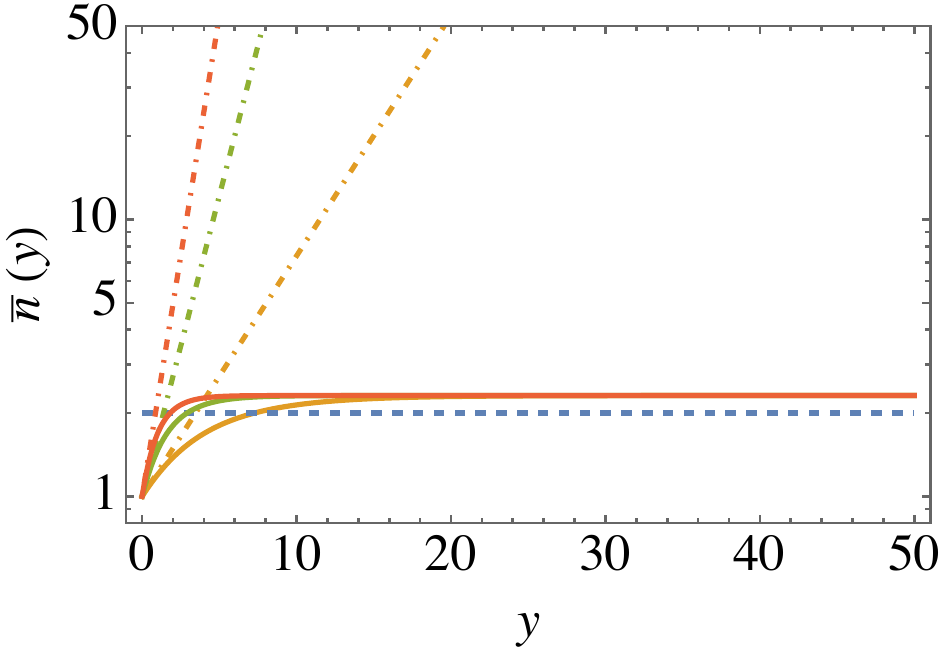} 
 \caption{Mean multiplicities of the $\beta$ cascade 
 (solid lines) as functions
 of $y$ for  $\alpha=0.2$ (orange), $\alpha=0.5$ (green) and $\alpha=0.8$ (red).
 Values of $r$:   0.05 (left panel) and 0.5 (right panel).
 For comparison dashed-dotted line corresponds to $\beta = 0$.
 Dashed blue line: $1/r$.}
  \label{fig:nbar_r}
\end{figure}

Next, we study the first two moments of the probability distribution generated by Eq.~(\ref{eq:EquationSat1}).
In Fig.~\ref{fig:nbar_r} we show the average multiplicity $\bar{n}(y)$ for $\alpha=0.2$, 0.5
and 0.8 and different values of $r$. For comparison we also show $\bar{n}(y)$ for $\beta = 0$,
{\em i.e.} for the $\alpha-$cascade without recombination.
One can see that for small $y$ multiplicity rises parallel to the
case $\beta=0$ and then saturates at the value, which depends only on $r$. 
For $r\le 0.2$ asymptotic value of the multiplicity 
$\bar{n}(y\rightarrow \infty)\simeq 1/r$. For larger $r$ multiplicities
scale to one common value, which is, however, larger than $1/r$. 
As is clear from 
Fig.~\ref{fig:nbar_r}, for fixed $r$ multiplicity saturates faster for larger $\alpha$.

\begin{figure}[h!]
  \centering
    \centering
 \includegraphics[width=7.5cm]{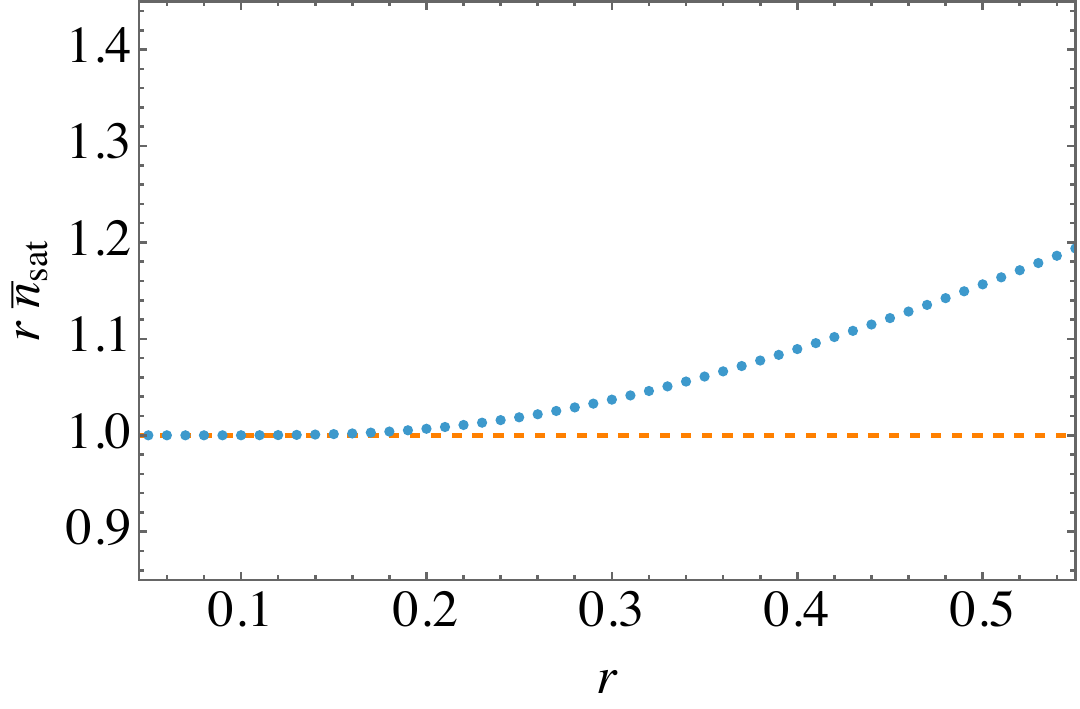}~
 \includegraphics[width=7.5cm]{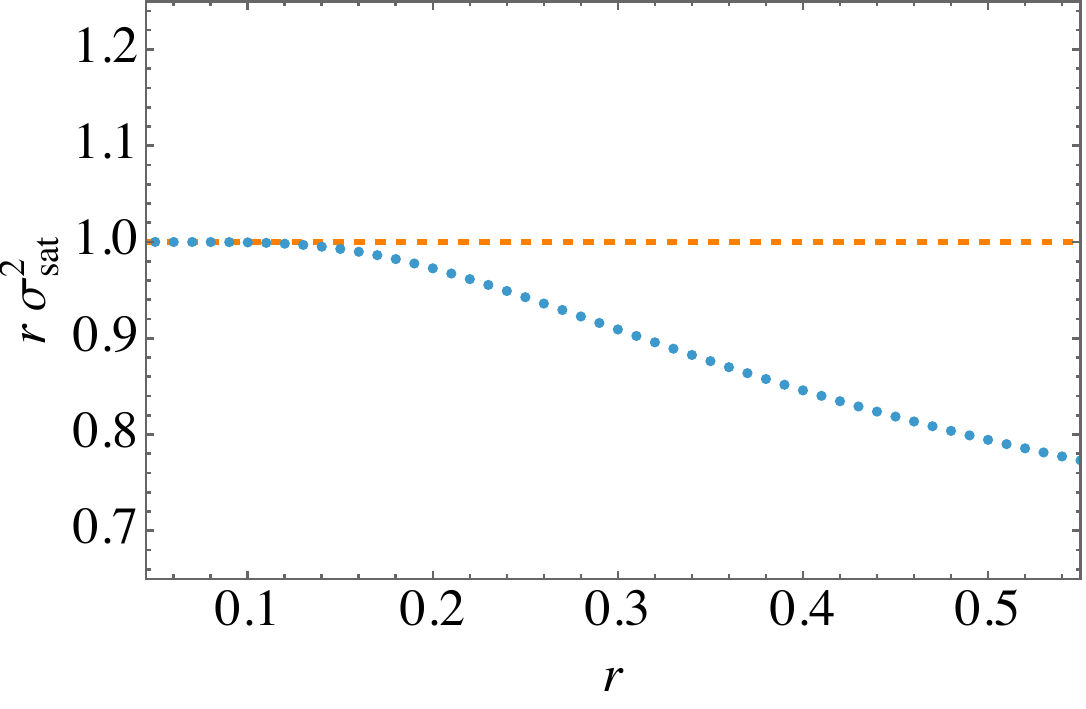}  
 \caption{Asymptotic values of $r \,\bar{n}_{\rm sat}$ and $r \, \sigma^2_{\rm sat} $ as functions of $r$. Dashed orange lines correspond
 to $1/r$ dependence discussed in the text.}
  \label{fig:nsigsat}
\end{figure}

As is clear from Fig.~\ref{fig:nbar_r} the saturation value of $\bar{n}$ fulfills the attractor scaling of 
Ref.~\cite{Blaizot:2006wp} $\bar{n}_{\rm sat}=1/r$ but only in the focal regime.
We illustrate this in the left panel Fig.~\ref{fig:nsigsat} where we show asymptotic values of $r \bar{n}$ as functions
of $r$.

\begin{figure}[h!]
  \centering
    \centering
     \includegraphics[width=7.5cm]{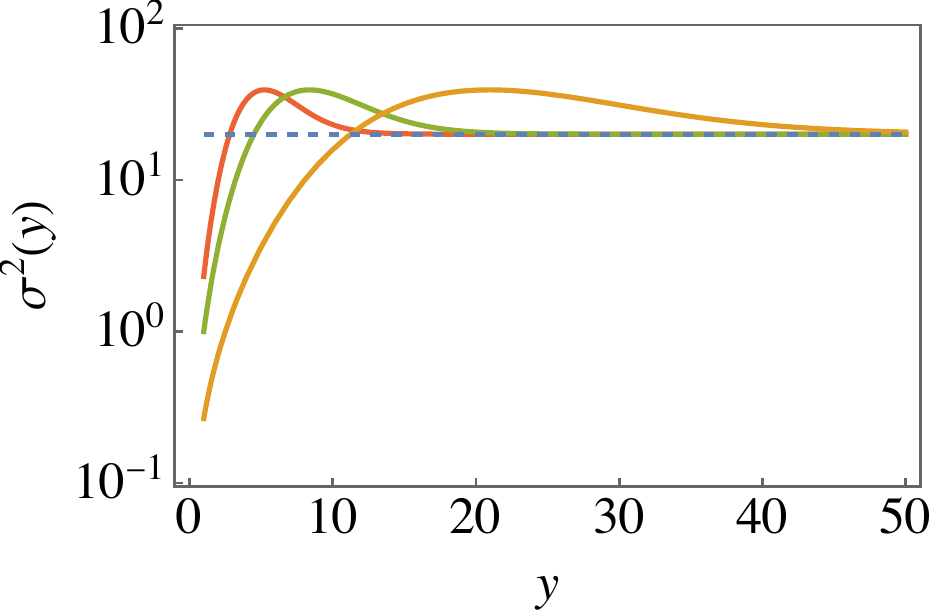}~ 
 \includegraphics[width=7.5cm]{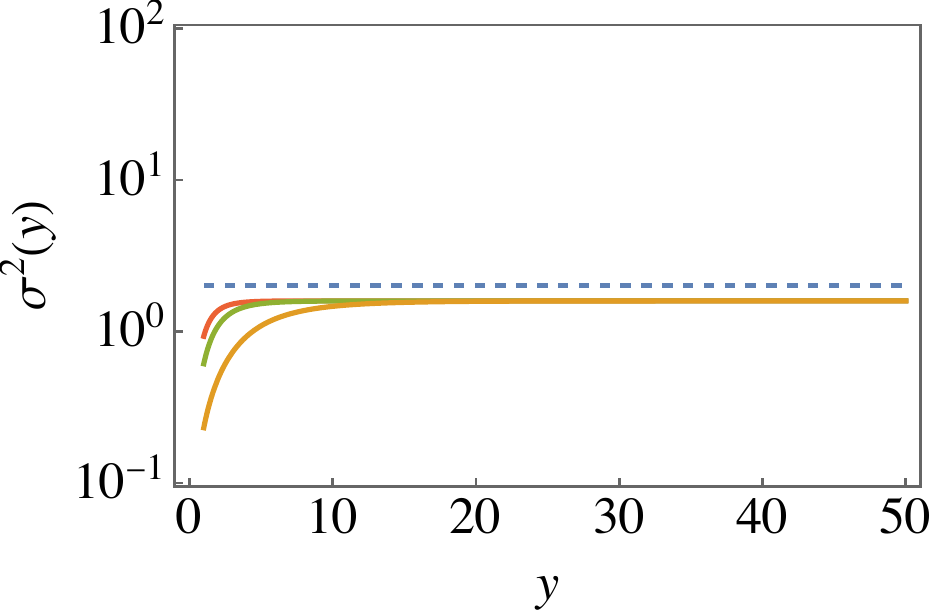}
 \caption{Variance of the $\beta$ cascade 
 (solid lines) as a function
 of $y$ for for $\alpha=0.2$ (orange), $\alpha=0.5$ (green) and $\alpha=0.8$ (red).
 Values of $r$: 0.05 (left panel) and 0.5 (right panel). Dashed blue line: $1/r$.}
  \label{fig:var_r}
\end{figure}

Next, we analyze variance $\sigma^2(y)$ for the same set of parameters as for $\bar{n}$ above.
 We see that in the focal regime (left panel of Fig.~\ref{fig:var_r}) $\sigma^2(y)$ has a maximum
which shifts to the left with increasing $\alpha$, and saturates for large rapidities.
For $r<0.2$ the asymptotic value is $1/r$, {\em i.e. }  $\sigma^2_{\rm sat}=\bar{n}_{\rm sat}=1/r$. 
However, as we will show below, this does not mean that the distribution is Poissonian.
 For larger values of $r$, in the parallel regime, the dependence on $r$
deviates from $1/r$. This is illustrated in the right panel of Fig.~\ref{fig:nsigsat}.

While the probability distribution for the $\alpha$ cascade is determined by the geometric distribution (\ref{eq:pneqPnG}) 
with the mean value of $\bar{n}_{\rm G}= \bar{n}-1$, it is reasonable to ask  if the
probability distributions given by generalized Eq.(\ref{eq:EquationSat1}), that accounts for
merging, can be described by the negative binomial distribution (\ref%
{eq:NBD}), such that it tends to (\ref{eq:pneqPnG}) for $\beta \rightarrow 0$. 
So we want to check if%
\begin{equation}
p_{n}(y)\overset{?}{=} p_n^{\rm NBD}(y) \equiv P_{n-1}^{\text{NBD}}(k(y),\bar{n}(y)-1),
\label{eq:conjecture}
\end{equation}%
which for $k=1$ indeed reduces to (\ref{eq:pneqPnG}). 

In order to fully determine the NBD we need to know parameter $k(y)$.
One can easily convince oneself that the variance of $p_n$ is equal to $\sigma^2_{\rm NBD}$.
Indeed
\begin{align}
\sigma^2& =\sum_{n=1}n^2 p_n - \bar{n}^2\\
&=\sum_{m=0}m^2 P_{m}^{\text{NBD}}(k,\bar{n}-1) - (\bar{n}-1)^2 \notag \\
&=\sigma^2_{\rm NBD} \, .
\end{align}
Hence we arrive at the formula for $k$
\begin{equation}
k=\frac{(\bar{n}-1)^2}{\sigma^2-(\bar{n}-1)} 
\label{eq:kascade}
\end{equation}
given in terms of $\bar{n}$ and variance of the probability
distribution $p_n$ of the $\beta$-cascade (\ref{eq:EquationSat1}).
One can see from Eq.~(\ref{eq:kascade}) that when $\sigma^2-\bar{n}=0$, which would normally
correspond to the Poisson distribution, the true distribution is still of the negative binomial type
with large but finite $k<\infty$. However, numerically it may be undistinguishable from the Poisson distribution with $k=\infty$.
This is a consequence of the fact that the $\beta-$cascade never fills $p_0$, {\em i.e.} a state without any particles.

\begin{figure}[h!]
  \centering
    \centering
 \includegraphics[width=7.5cm]{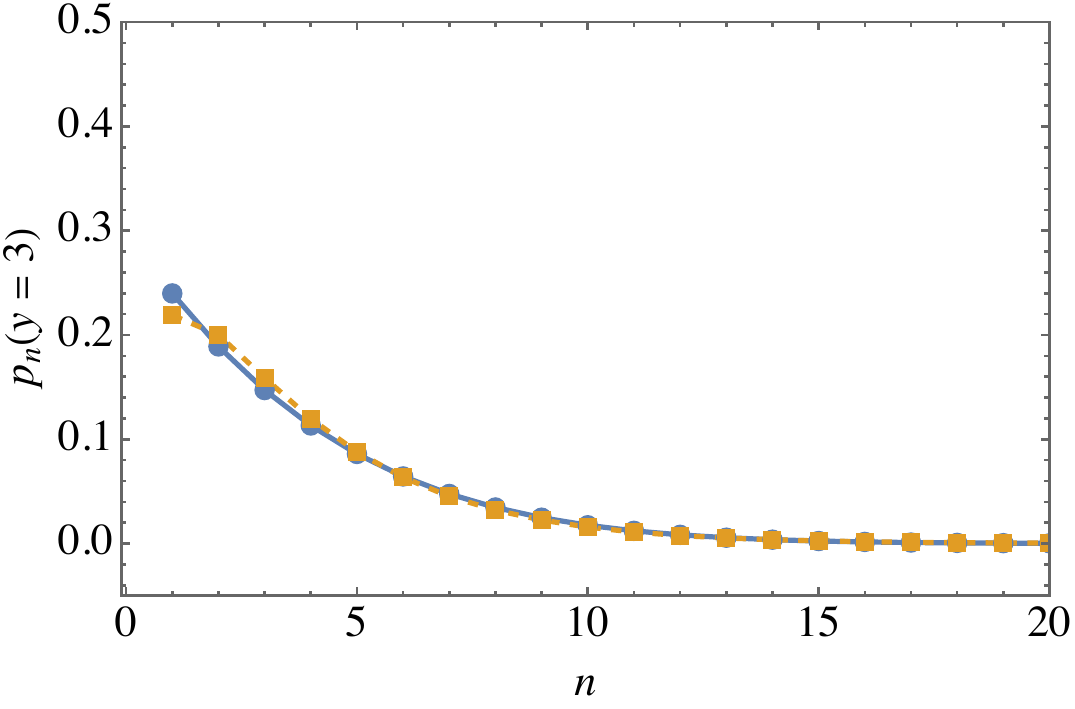}~
  \includegraphics[width=7.5cm]{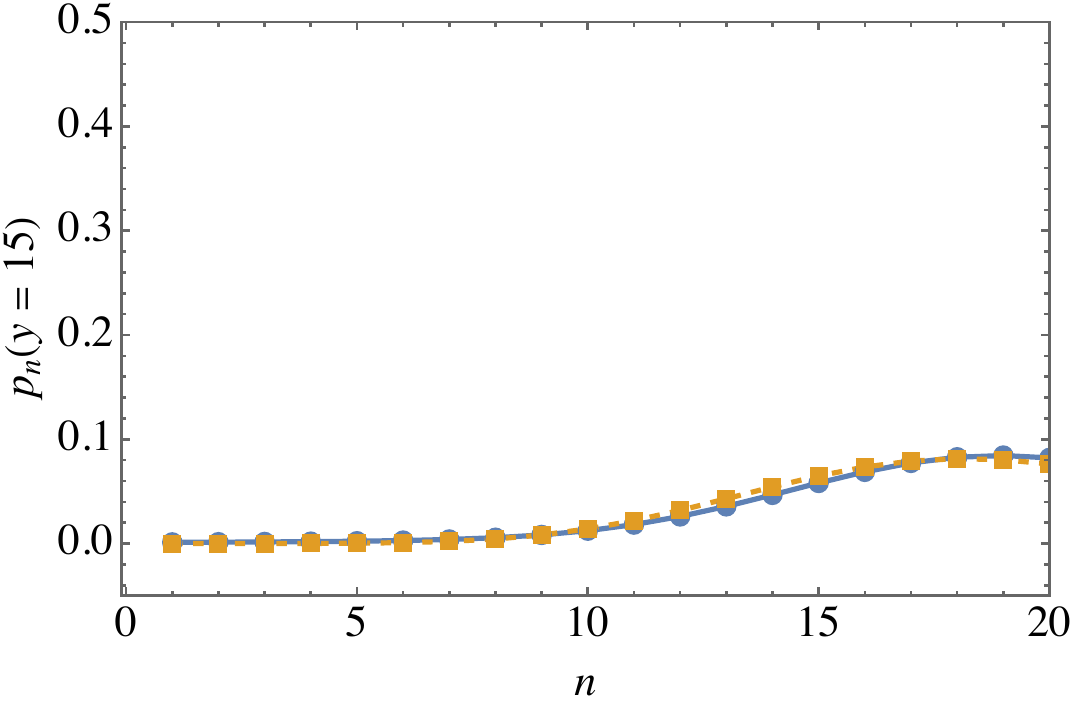}\\
  \includegraphics[width=7.5cm]{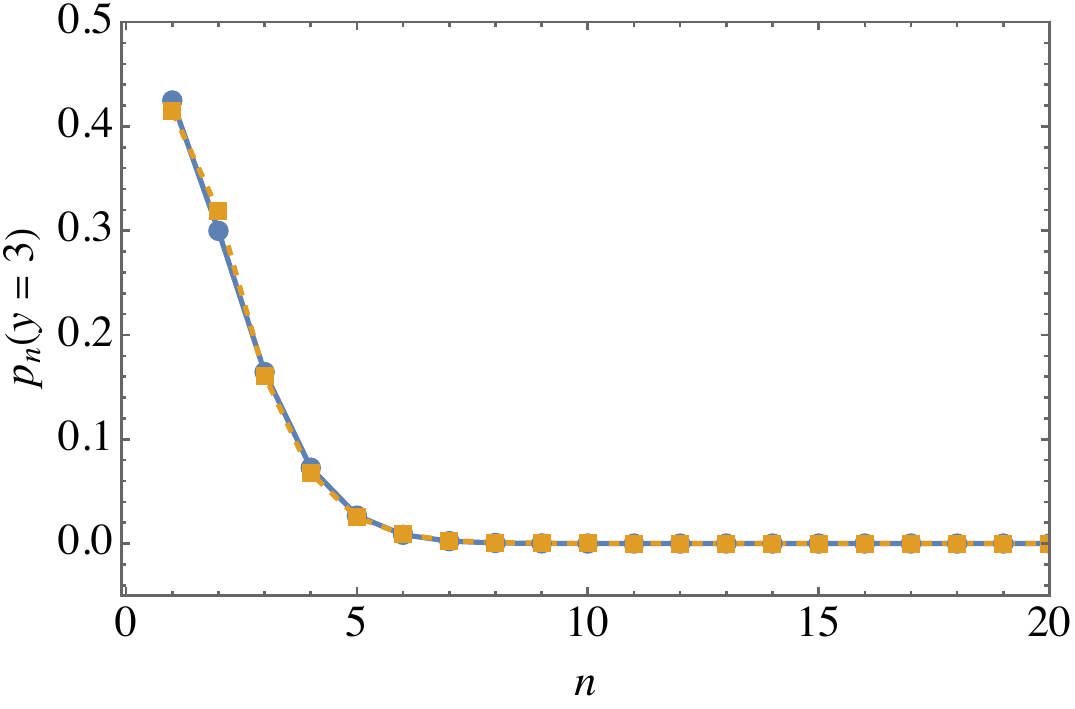}~
 \includegraphics[width=7.5cm]{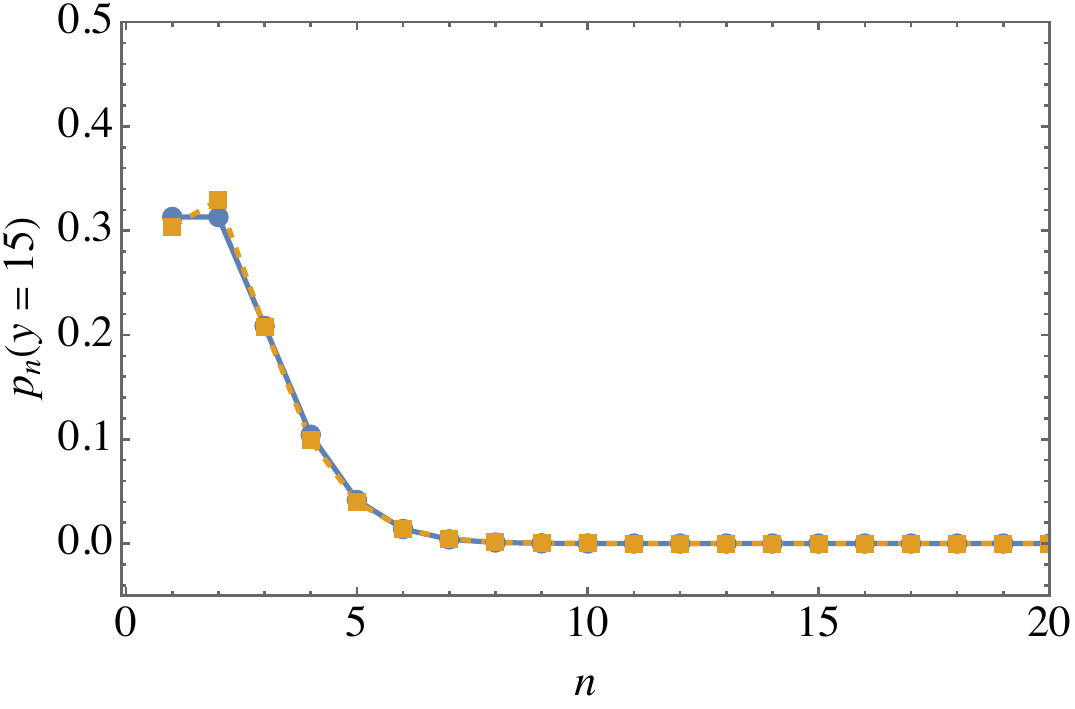} 
 \caption{Probability distributions $p_n(y)$ for the $\beta$-cascade (blue circles)
 with $\alpha=0.5$
 and the corresponding NBD (orange squares) for selected values of $y=3$ and 15.
 In the upper row we have plotted cascade for $r=0.05$, whereas the lower row
 corresponds to $r=0.5$. Lines are drawn to guide the eyes.}
  \label{fig:pnBKy}
\end{figure}

To verify the conjecture (\ref{eq:conjecture}) we have evolved Eq.~(\ref{eq:EquationSat1}) for $\alpha=0.5$ and $r=0.05$ (focal regime),
and $r=0.5$ (parallel regime). The results are shown in Fig.~\ref{fig:pnBKy} as blue circles. For each case we have
computed the corresponding $\bar{n}$ and $k$, and compared $p_n(y)$ and $p_n^{\rm NBD}(y)$ shown in Fig.~\ref{fig:pnBKy} as orange squares. 
Looking at Fig.~\ref{fig:pnBKy} one concludes that the NBD is almost indistinguishable from the numerical multiplicity
distribution of the $\beta$-cascade both in the focal and parallel regimes. 
We have checked that the same behavior is observed for other values of the cascade parameters. To test how well
NBD describes the $\beta$-cascade we have also
computed the third moment. The difference between the true value and the NBD value is indeed very small.

\begin{figure}[h!]
  \centering
    \centering
  \includegraphics[width=7.5cm]{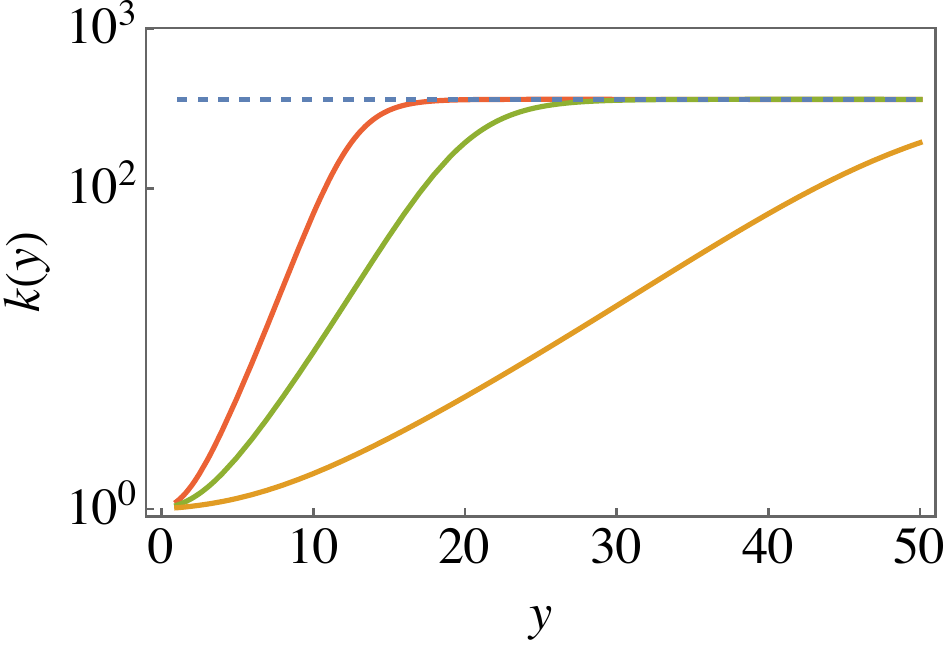} ~
 \includegraphics[width=7.5cm]{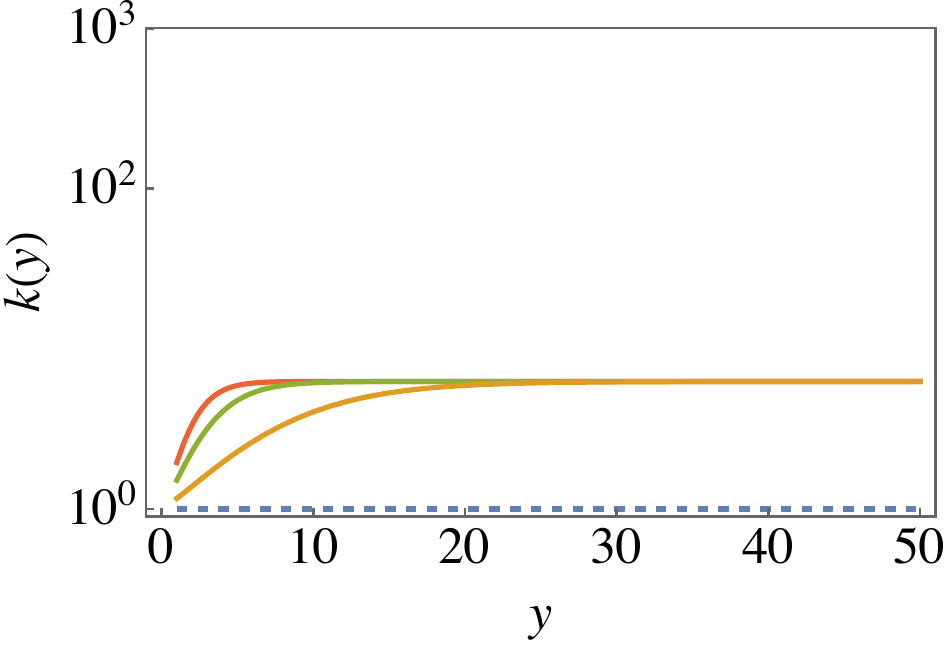}
 \caption{Parameter $k$ of the NBD for the $\beta$-cascade 
 (solid lines) as a function
 of $y$ for $\alpha=0.2$ (orange), $\alpha=0.5$ (green) and $\alpha=0.8$ (red).
 Values of $r$: 0.05 (left) and 0.5 (right).
 Dashed blue line corresponds to the saturation value $k_{\rm sat}=(1/r-1)^2$.}
  \label{fig:kNBD_r}
\end{figure}

\begin{figure}[h!]
  \centering
    \centering
 \includegraphics[width=7.5cm]{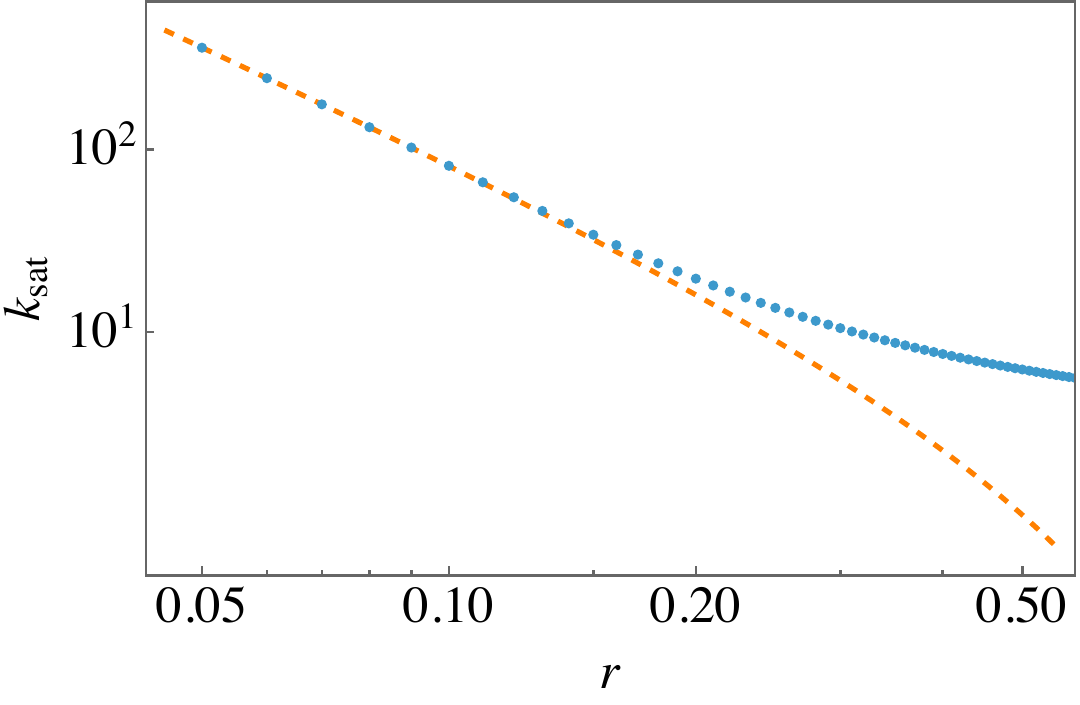}
 \caption{Asymptotic values of $k_{\rm sat}$ a function of $r$. Dashed orange line corresponds
 to $(1/r-1)^2$.}
  \label{fig:ksat}
\end{figure}

An important and interesting question is the energy dependence of the NBD parameter $k$ (\ref{eq:kascade}).
From our previous discussion on saturation of $\bar{n}$ and $\sigma^2$, see Fig.~\ref{fig:nsigsat}, we expect that $k$ should
saturate for large $y$ at $k_{\rm sat}=(1/r-1)^2$, at least in the focal region. To this end in Fig.~\ref{fig:kNBD_r}
we plot parameter $k$ as a function of $y$ for different couplings $\alpha$ and two values of $r$.
We see that for small $r$ (focal regime) parameter $k$ saturates at $k_{\rm sat}$
defined above, whereas for larger $r$ (parallel regime) the saturation sill takes place but at a different
value.
Saturation is faster for larger $\alpha$. The $r$ dependence
of $k_{\rm sat}$ is shown in  Fig.~\ref{fig:ksat}. Interestingly, independently of $\alpha$ and $r$
the asymptotic value of $k$ is always finite, and therefore the probability distribution does not have
a Poissonian limit for large $y$. For small $r$ parameter $k$ saturates at $k_{\rm sat}$ that is growing
with $r \rightarrow 0$.

Let us summarize our findings for the $\beta$-cascade. The cascade shows a scaling behavior for large $y$:
average multiplicity, variance and the NBD parameter $k$ saturate for large $y$ at a value which is
depends only on $r$. Saturation is achieved faster for larger $\alpha$. In the focal regime ($r < 0.2$) the scaling
behavior is very simple
\begin{align}
\bar{n} &\rightarrow \bar{n}_{\rm sat}=\frac{1}{r}\, , \notag \\
\sigma^2& \rightarrow \sigma^2_{\rm sat}=\frac{1}{r}\, , \notag \\
k     &\rightarrow k_{\rm sat}=\Big( \frac{1}{r}-1 \Big)^2 \, .
\label{eq:satscaling}
\end{align}

One should note that the above scaling laws have to break at $r \rightarrow 0$.
We know that for the BFKL-like cascade ($r=0$) $\bar{n}$ does not saturate 
and grows with $y$ to infinity (\ref{eq:nbarBFKL}), which is consistent with the $1/r$ behavior.
However, in the case of geometric distribution (\ref{eq:pdistr}) the variance $\sigma^2_{\rm G}=\bar{n}_{\rm G}(\bar{n}_{\rm G}+1)=\bar{n}(\bar{n}-1)$ 
does not scale as $1/r$
but rather as $1/r^2$
and, as a consequence, $k=1$. In contrast, for the $\beta$-cascade 
$k$ tends to $\infty$ for small $r$, Fig.~\ref{fig:ksat}, therefore,
the transition from the NBD to the geometric distribution with $k=1$ is discontinuous.

\subsection{Cascade $\gamma$}

\begin{figure}[h!]
  \centering
    \centering
 \includegraphics[width=7.5cm]{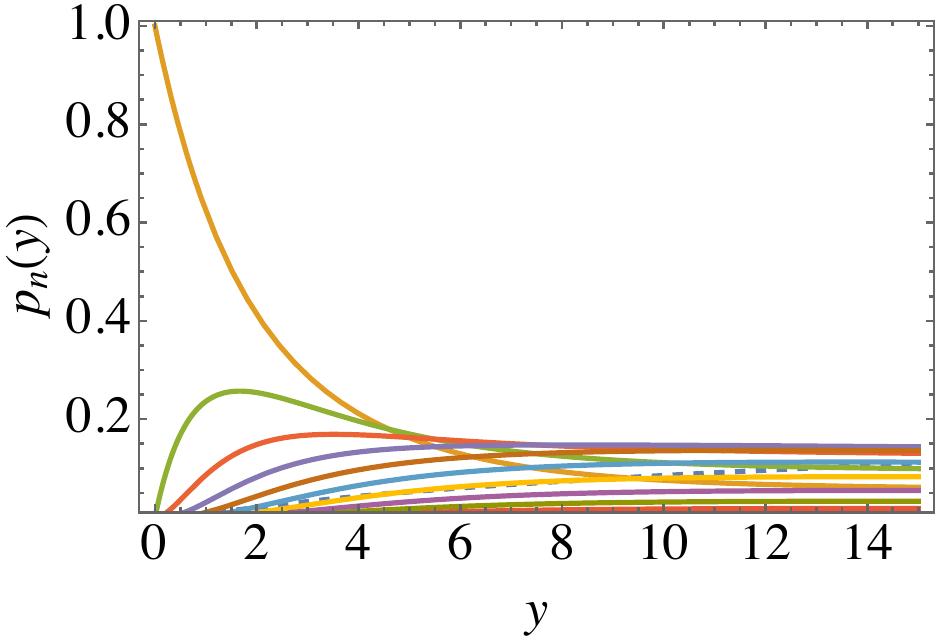}~
 \includegraphics[width=7.5cm]{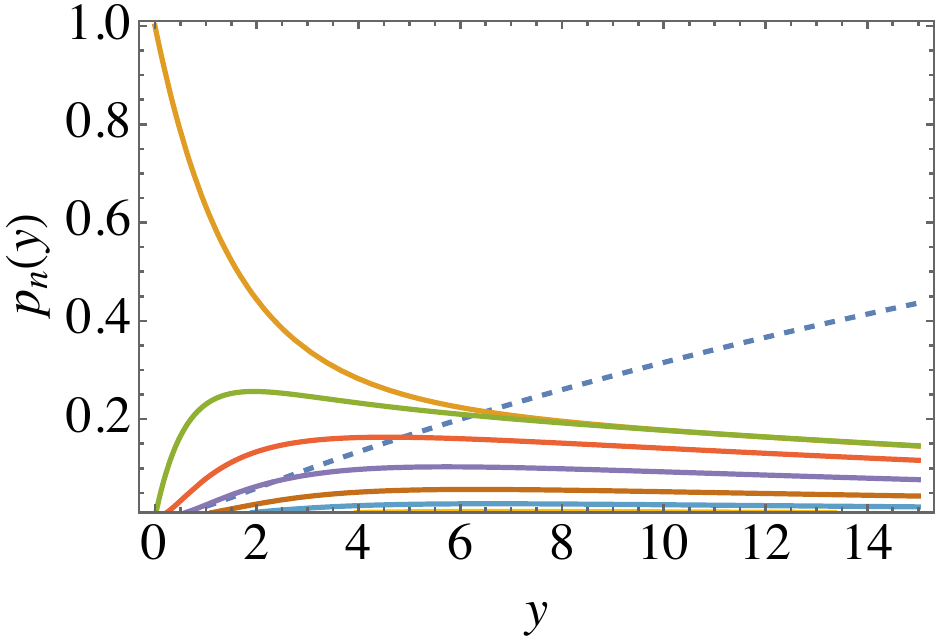}
\caption{Probabilities $p_n(y)$ ($n=0,\ldots ,10$) for the $\gamma$ branching (\ref{eq:eqsat2})
 for $\alpha=0.5,~r=0.1$ and $s=0.5$ (left),  
 and $s=1.5$ (right). Probability $p_0$ is shown as a blue dashed line.
The corresponding $\beta$-cascade is shown in the left panel of Fig.~\ref{fig:pnyr}.}
  \label{fig:pngyrs}
\end{figure}

Now we turn to Eq.~(\ref{eq:eqsat2}) describing the  cascade with recombination and with additional
$\gamma$-term ($\gamma$-cascade), which we parametrize as
\begin{equation}
   \gamma=s \beta \, . 
\end{equation}
The $\gamma$-term is responsible for the disappearance of particles
during evolution. An immediate consequence of this process is an 
increase of $p_0$ with $y$, which was identically zero in the case of cascade (\ref{eq:EquationSat1}).
Moreover, the $\gamma$ term is {\em defocusing} the probability distribution. This is shown in Fig.~\ref{fig:pngyrs}
where the $\gamma$ term with $s=0.5$ and 1.5 is added to the $\beta$-cascade shown
in the left panel of Fig.~\ref{fig:pnyr}.

\begin{figure}[h!]
  \centering
    \centering
 \includegraphics[width=7.5cm]{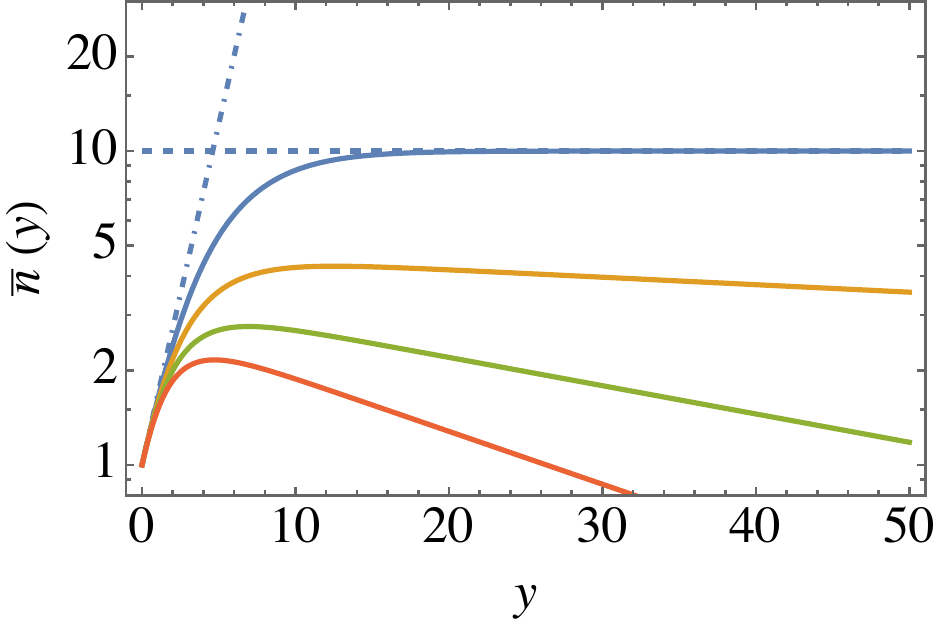}~
 \includegraphics[width=7.5cm]{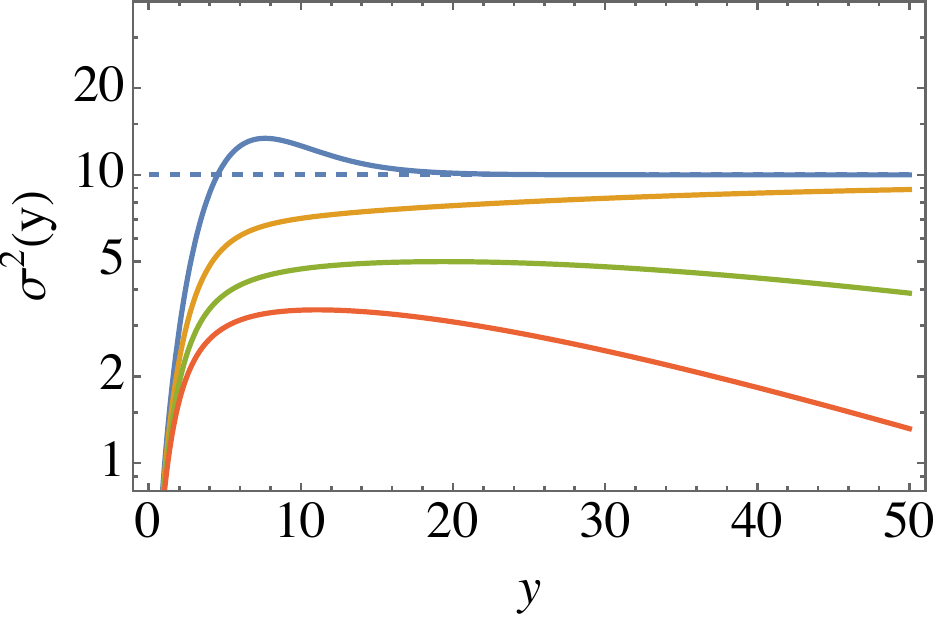}
\caption{Mean multiplicity (left) and variance (right) for the $\gamma$ branching (\ref{eq:eqsat2})
 for $\alpha=0.5,~r=0.1$ and $s=0,~0.5,~1,~1.5$ (from top to bottom). 
 Horizontal dashed lines show saturation values for the corresponding $\beta$-casccade.
 In the left panel mean value for the corresponding $\alpha$-cascade is shown with a dashed-dotted line. }
  \label{fig:nbarvargys}
\end{figure}

In Fig.~\ref{fig:nbarvargys} we show the mean value and variance for the $\gamma$-cascade
with $\alpha=0.5$ and $r=0.1$ for different values of $s$. For comparison 
$\bar{n}$ and $\sigma^2$ for the corresponding $\beta$-cascade are shown in blue (upper curves)
together with the asymptotic values shown by dashed lines. Mean multiplicity for the
$\alpha$-cascade is shown by dashed-dotted line. We see that both average multiplicity
and variance do not saturate and start decreasing for large $y$. The decrease is faster for larger $s$.
This pattern is general and does not depend on $\alpha$ and $r$. For larger $s$ the decrease
of $\bar{n}$ and $\sigma^2$ is getting stronger.

In the case of $\gamma$-cascade the identification (\ref{eq:conjecture})
breaks down. However, it is conceivable that after subtracting $p_0$, 
the remainder does still behave like an NBD. Subtracting $p_0$ from a normalized
$\gamma$-cascade distribution makes the rest unnormalized. To correct for this,
we define a new distribution
\begin{equation}
    p_n \rightarrow {p}_{\gamma n}=\frac{p_n}{1-p_0},~~~n=1,2,\ldots \, ,
    \label{eq:tildep}
\end{equation}
where $p_n$'s are numerical solutions of Eq.~(\ref{eq:eqsat2}). Note that
the normalized
${p}_{\gamma n}$ distribution has a different mean than $p_n$ previously denoted
by $\bar{n}$ (note that $n=0$ does not contribute to the mean value)
\begin{equation}
    \bar{n}_{\gamma}=\frac{\bar{n}}{1-p_0} \, .
\end{equation}

Analogously, the variance reads
\begin{equation}
   \sigma^2_{\gamma}=\sum_{n=1}n^2{p}_{\gamma n}-\bar{n}_{\gamma}^2=
   \frac{1}{(1-p_0)^2}\Big( (1-p_0)\sum_{n=1}n^2 p_n -\bar{n}^2 \Big)
   \equiv \frac{1}{(1-p_0)^2}\tilde{\sigma}^2_{\gamma} \, .
\end{equation}
With this notation the pertinent $k_{\gamma}$ parameter is given by
\begin{equation}
    k_{\gamma}=\frac{(\bar{n}-(1-p_0))^2}{\tilde{\sigma}^2_{\gamma}-(1-p_0)(\bar{n}-(1-p_0))} \, .
\end{equation}
Hence, we assume that for $n \geq 1$
\begin{equation}
{p}_{\gamma n}(y)\overset{?}{=} {p}_{\gamma n}^{\rm NBD}(y) \equiv P_{n-1}^{\text{NBD}}(k_{\gamma},\bar{n}_{\gamma}-1 ) \, .
\end{equation}
Therefore the  probability distribution is assumed to be
\begin{equation}
 p_n=  \{ p_0,(1-p_0)\,{p}_{\gamma n}^{\rm NBD} \} \, .
    \label{eq:all4gamma}
\end{equation}
Recall that $\bar{n}$ is the multiplicity of the $\gamma$-cascade, whereas $\bar{n}_{\gamma}$ 
is the multiplicity of the properly normalized remainder, after
discarding $p_0$. Likewise, $k_{\gamma}$ is the $k$ parameter of the NBD
approximating the remainder. Therefore, distribution (\ref{eq:all4gamma})
depends on three parameters $p_0$, $\bar{n}$ and $k_{\gamma}$.

\begin{figure}[h!]
  \centering
    \centering
 \includegraphics[width=7.5cm]{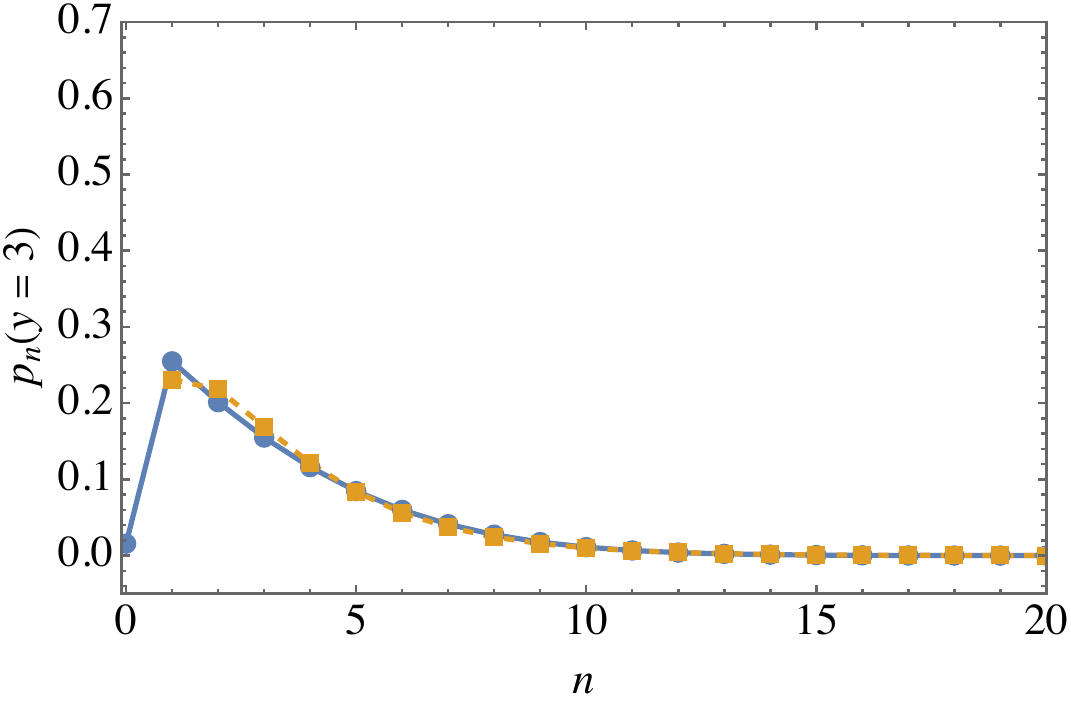}~
  \includegraphics[width=7.5cm]{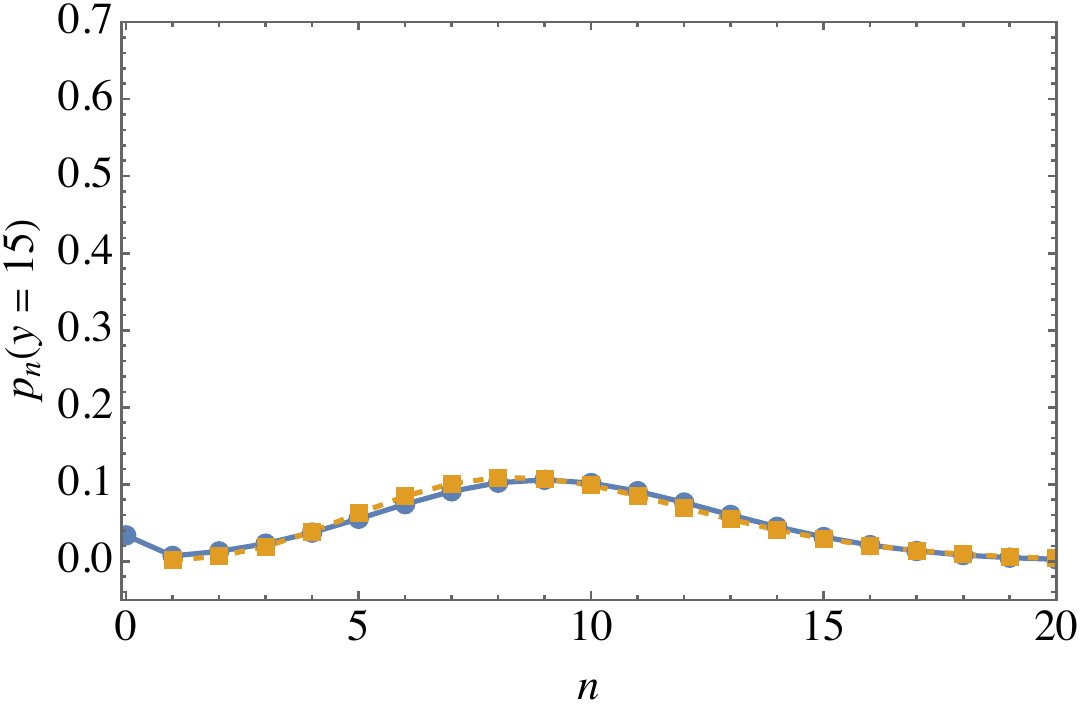}\\
  \includegraphics[width=7.5cm]{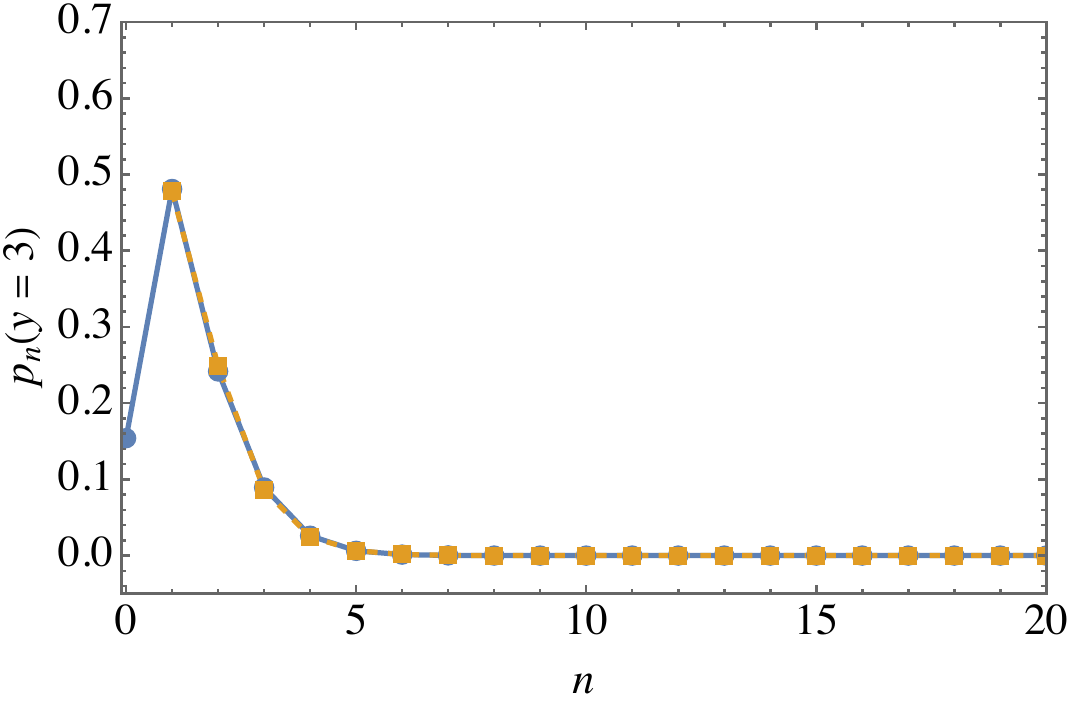}~
 \includegraphics[width=7.5cm]{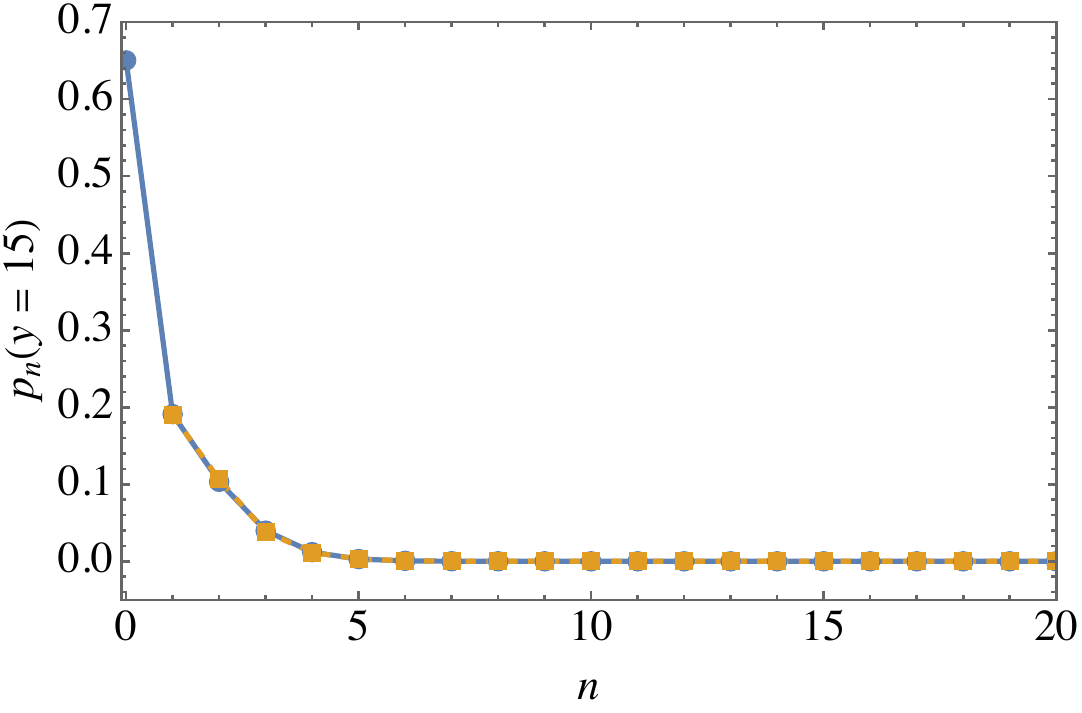} 
 \caption{Probability distributions $p_n(y)$ for the $\gamma$-cascade (blue circles)
  with $\alpha=0.5$ and $s=0.5$
 and the corresponding NBD (orange squares) for selected values of $y=3$ and 15.
 In the upper row we have plotted cascade for $r=0.05$, whereas the lower row
 corresponds to $r=0.5$. Lines are drawn to guide the eyes.}
  \label{fig:pngy}
\end{figure}

\begin{figure}[h!]
  \centering
    \centering
 \includegraphics[width=7.5cm]{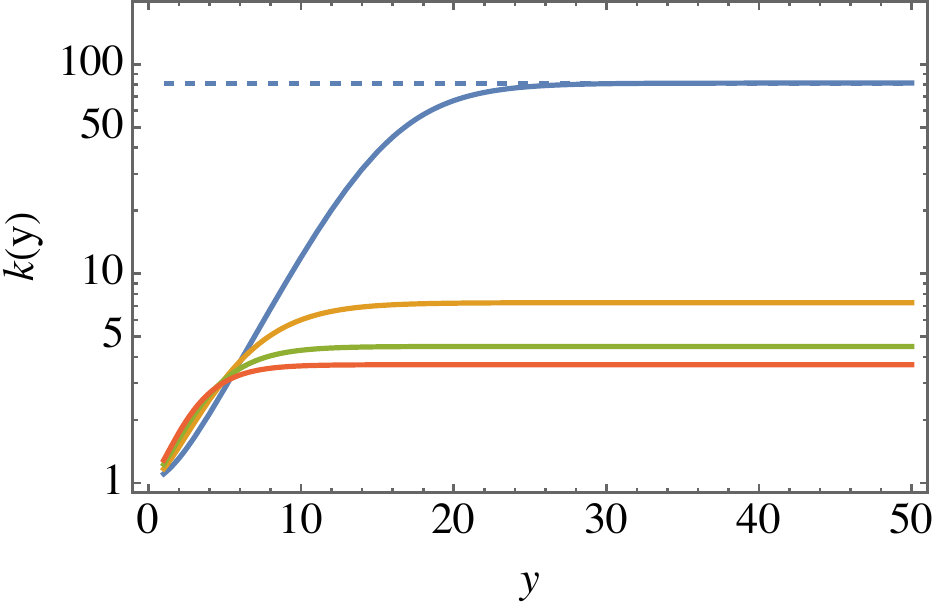}~
 \includegraphics[width=7.5cm]{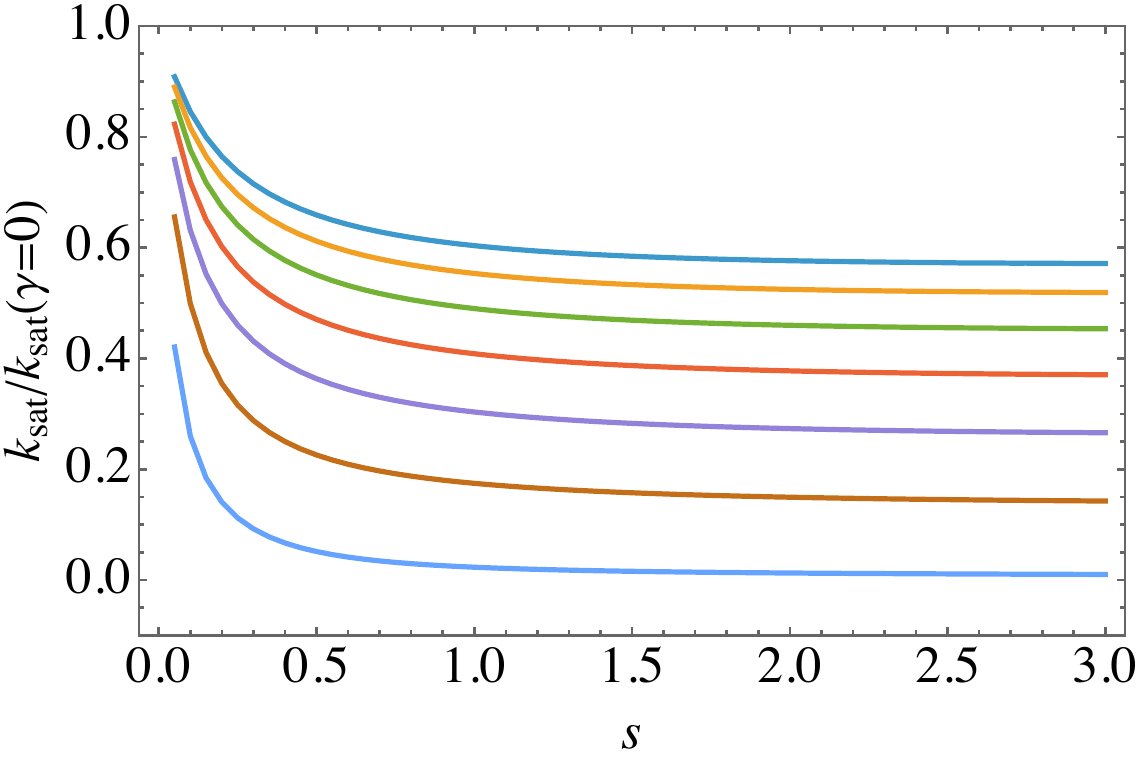}
\caption{ NBD parameter $k$ for the $\gamma$-cascade. In left panel $k$ is plotted
as a function of $y$ for $\alpha=0.5,~r=0.1$ and different values of $s=0,0.5,1,1.5$ (from top to bottom).
In the right panel ratios of saturation values of $k_{\gamma\, {\rm sat}}/k_{\rm sat}$ are plotted as functions of
$s$ for different values of $r=0.7,0.6,0.5,0.4,0.2,0.05$ (from top to bottom).}
  \label{fig:kgamma}
\end{figure}

In Fig.~\ref{fig:pngy} we plot the probability distributions $p_n(y)$ for the $\gamma$-cascade
with $\alpha=0.5$ and $s=0.5$,
 and the corresponding NBD  for selected values of $y=3$ and 15. We can see that 
 probability $p_0$ grows with $y$ and the  growth is faster for larger $r$. The reminder
 is very well approximated by the NBD (\ref{eq:all4gamma}). In the left panel of Fig.~\ref{fig:kgamma}
 we plot rapidity dependence of $k_{\gamma}$ for different values of $s$. Unlike $\bar{n}$ and
 $\sigma^2$, which decrease with $y$, see Fig. \ref{fig:nbarvargys}, parameter $k_{\gamma}$ saturates 
 for large $y$ at a value smaller than for the $\beta$-cascade,
 see the right panel of Fig.~\ref{fig:kgamma}. Therefore, an interesting picture arises: in the case
 of the $\gamma$-cascade  $p_0 \rightarrow 1$ asymptotically, and the remaining probabilities tend to 0,
 preserving, however, the shape of the NBD with constant $k_{\gamma}$. The driving force behind nullification of
 $p_{n>0}$ is vanishing $\bar{n}_{\gamma}$ at constant $k_{\gamma}$.


\section{Entropy and other quantum measures}
\label{sec:EQI}

In this section we discuss the behavior of entropy and other quantum measures for the
cascades studied in the previous sections. 
The von Neuman entropy that follows from integrating  out all degrees of freedom except rapidity \cite{Kharzeev:2017qzs,Liu:2022hto} 
 reads
\begin{equation}
    S(y) = -\sum_n p_n(y) \ln ( p_n(y)) \, ,
    \label{eq:Sdef}
    \end{equation}
where $p_n(y)$ are interpreted as probabilities for a given number of partons at rapidity $y$. 
 It measures how much information has been lost due to integrating out quarks and averaging over transverse 
and  color degrees of freedom, as evolution in rapidity progresses. 
In the complete 3+1 D dipole model the entropy depends also on the  hard scale of the measurement process, while in the 1 D model 
it might be introduced  by promoting the partonic multiplicity to parton density function and letting it to depend on the hard scale. 

In the case of the $\alpha$-cascade, where the probability distribution given by (\ref{eq:pdistr}), the entropy can be
calculated analytically \cite{Kharzeev:2017qzs}
\begin{equation}
    S=\bar{n}\ln \bar{n}-(\bar{n}-1) \ln (\bar{n}-1) \, ,
    \label{eq:SDima}
\end{equation}
where $\bar{n}(y)$ is given by (\ref{eq:nbarBFKL}). For large $y$  entropy $S\rightarrow \ln \bar{n}=\alpha y$ and grows linearly
with rapidity. This is a sign of maximal entanglement \cite{Hentschinski:2022evidence,Hentschinski:2023maxent}.

\begin{figure}[h!]
  \centering
    \centering
 \includegraphics[width=7.0cm]{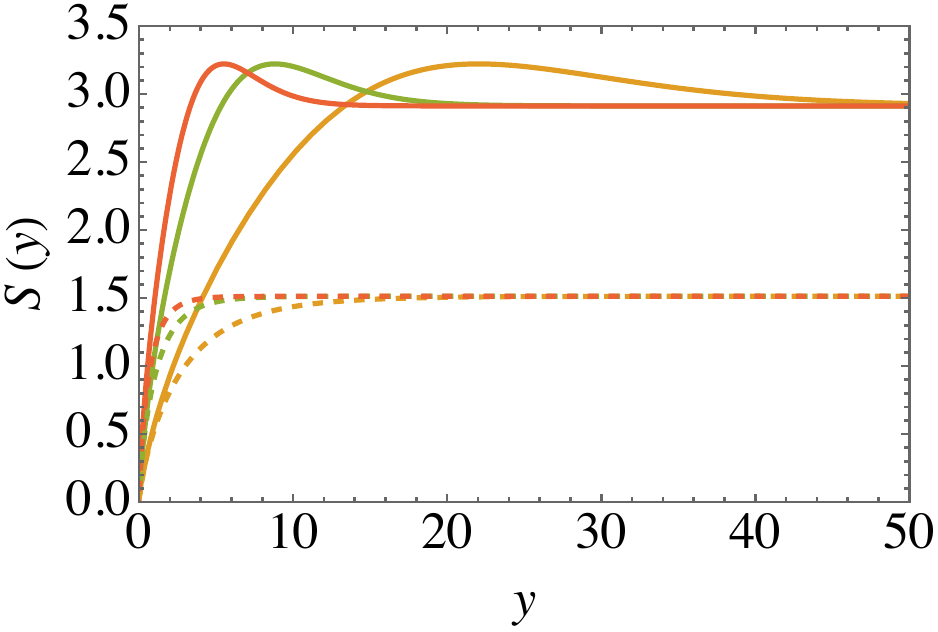}~
 \includegraphics[width=7.0cm]{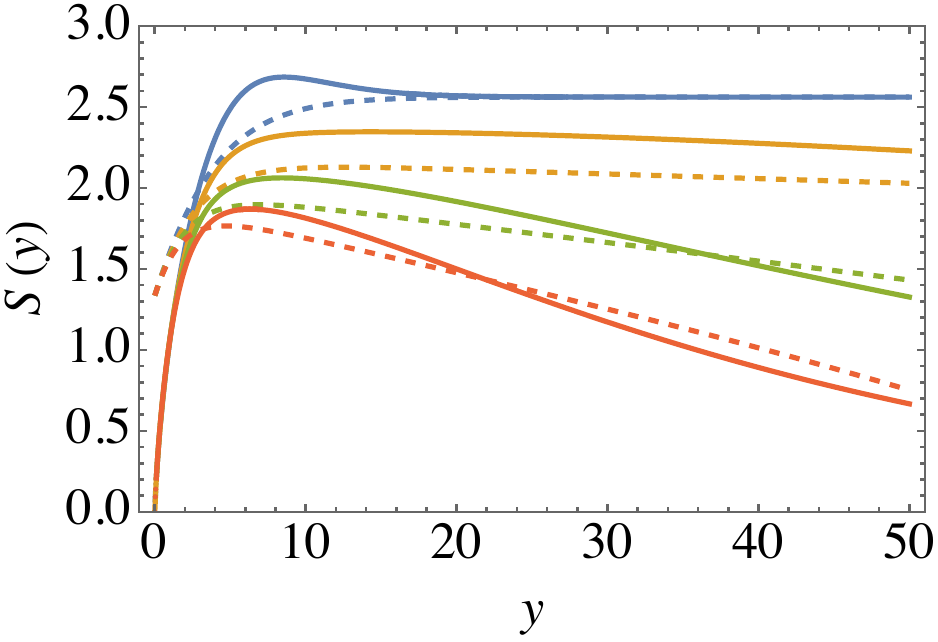}\\
\caption{Entropy of the $\beta$ (left)  and $\gamma$ (right) cascade   
as functions of  $y$ for $r=0.05$ (upper left) and $r=0.5$ (lower left)
for three values of $\alpha$: 0.2 (orange), 0.5 (green) and 0.8 (red). In the right panel the $\beta$ cascade (upper blue line) for $\alpha=0.5$ and $r=0.1$ is compared with the
$\gamma$ cascade with $s=0.5,1,1.5$ (top to bottom).
Approximation (\ref{eq:SHatta}) is shown as dashed lines.}
\label{fig:Sy}
\end{figure}

Recombination processes slow down the probability evolution and -- as we have shown in Sect.~\ref{sec:beta} -- the multiplicity
and the NBD parameter $k$ saturate. We expect that also entropy saturates.
In the left panel of Fig. \ref{fig:Sy} we plot entropy 
generated in $\beta$ branching process
as a function
of rapidity $y$ for small (upper solid lines) and large (lower dashed lines) value of parameter $r$.
For small $r=0.05$ the system is in the focal regime and the maximum of entropy
occurs at some finite $y$ corresponding to the probabilities crossing shown in
the left panel of Fig.~\ref{fig:pnyr}. 
This behavior has been already observed in Ref.~\cite{Hagiwara:2017uaz}.
After the bump the entropy saturates at some
finite value.
In the left panel Fig. \ref{fig:Sy} we also show the entropy
for the cascade with $r=0.5$ in the parallel regime (dashed lines). We see that there is no bump
and the entropy monotonically reaches the saturation value. In both cases the
approach to saturation is faster for larger $\alpha$. Since recombination processes lead to saturation, 
we conclude that saturation prevents the system from reaching maximal entanglement.

It is interesting to compare our results for entropy with approximate formulas relating entropy to
the average multiplicity $\bar{n}$. To this end we use (\ref{eq:SDima}) and an approximate relation between entropy
and $\bar{n}$ for the Poisson distribution \cite{Hagiwara:2017uaz}
\begin{equation}
    S=\frac{1}{2}\ln(2\pi e \,\bar{n})-\frac{1}{12\,\bar{n}} + \ldots \, .
    \label{eq:SHatta}
\end{equation}
We expect (\ref{eq:SHatta}) to work well for the $\beta$-cascade where the probability distribution
is very close to the Poisson distribution. That this is indeed the case can be seen in the right panel
of Fig.~\ref{fig:Sy} where (\ref{eq:SHatta}) is shown as dashed lines. Approximation (\ref{eq:SDima}) is
not shown, as it misses the numerical results.
On the contrary, one can see that (\ref{eq:SHatta}) very well reproduces 
the $\beta$-cascade for large $y$ (upper blue lines), and to a lesser accuracy the $\gamma$-cascade.
The latter is to be expected since the probability distribution (including $p_0$) is in this case neither NBD nor Poissonian.
Nevertheless, we see that the decrease of entropy with $y$ is driven by a decrease of $\bar{n}$.

In addition to entropy, solutions to the cascade equations can be used to calculate various  quantum information (QI) measures. 
In the paper \cite{Caputa:2024xkp} it has been demonstrated that the equation for the BFKL-like cascade can be directly related 
to the Schr{\"o}dinger equation for a state evolved with a boost operator having SL(2,R) symmetry. 
This relation established a link between QI measures and the 1 D dipole model.
In particular in the mean number of dipoles $\bar{n}$, called in the QI context a complexity, measures how the underlying 
quantum state spreads in the Hilbert space \cite{Baiguera:2025dkc}. Quantum complexity is here of lesser interest to us,
because parton multiplicities are only indirectly related to measured hadronic multiplicities.
The variance has been already introduced earlier, and it quantifies the fluctuations around the average. Here we use normalized variance 
defined as follows
\begin{equation}
    \delta^2=\frac{\langle n^2\rangle-\langle n\rangle^2}{\langle n\rangle^2} \, .
\end{equation}
Finally, the quantity that measures the purity, {\em i.e.} how the system departures from a pure state as evolution progresses,
is defined as
\begin{equation}
    \gamma=\sum_n p_n^2 \, .
    \label{eq:purdef}
\end{equation}
This quantity is of special interest in this paper as it is bounded: for the maximally mixed state the purity is zero and for the pure state it is equal to one. 

\begin{figure}[t!]
  \centering
  \begin{minipage}[b]{0.45\textwidth}
    \centering
    \includegraphics[width=\linewidth]{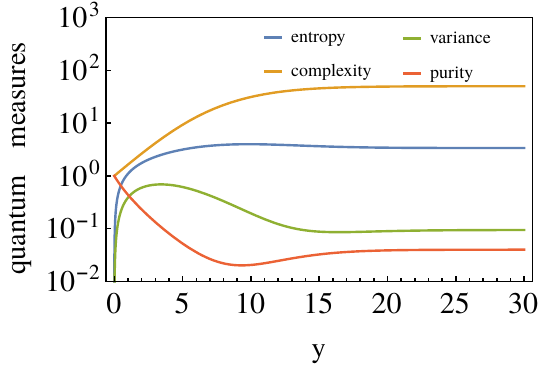}
  \end{minipage}
 \begin{minipage}[b]{0.45\textwidth}
    \centering
    \includegraphics[width=\linewidth]{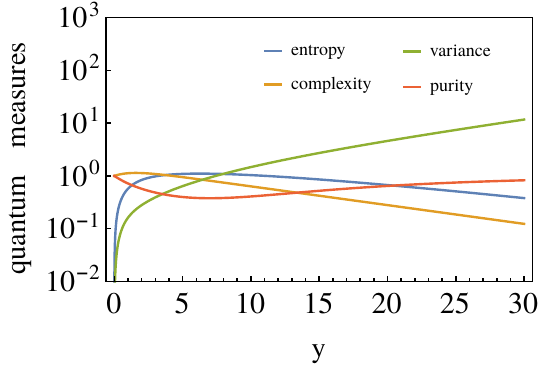}
  \end{minipage}\hfill
  \caption{ In the left panel: quantum measures as obtained from solutions of $\beta$-cascade Eq.~(\ref{eq:EquationSat1}). 
  In the right panel: quantum measures as obtained from solutions of $\gamma$-cascade Eq.~(\ref{eq:eqsat2}).}
  \label{fig:results22}
\end{figure}

In the following, we apply the Quantum Measures to the $\beta$- and $\gamma$-cascades, which is illustrated in Fig.~\ref{fig:results22}. 
The QI measures were introduced to study dipole cascades in \cite{Caputa:2024xkp} and have already been applied to the $\beta$-cascade. 
However, it is instructive to contrast the results obtained for 
the $\beta$-cascade with those for the $\gamma$-cascade. We see that at the beginning of the evolution, the QI measures are similar, 
which indicates the dominance of the linear or splitting region. However, as nonlinear corrections and the transition 
to the vacuum become significant, the quantities differ substantially. In particular, we see that the entropy of 
the $\gamma$-cascade starts to drop, similarly 
to the complexity, whereas in the case of the $\beta$-cascade, they saturate. We also observe that as the entropy increases, 
the purity of the $\gamma$-cascade, after an initial drop, starts to increase, indicating that entanglement in the system decreases.
On the other hand, the growing variance of the $\gamma$-cascade indicates  large fluctuations, which is consistent with the growing 
probability of transitions to the vacuum.


\section{LHCb hadronic entropy and purity vs. cascade models}
\label{sec:LHCb}

\begin{figure}[t!]
  \centering
  \begin{minipage}[b]{0.45\textwidth}
    \centering
    \includegraphics[width=\linewidth]{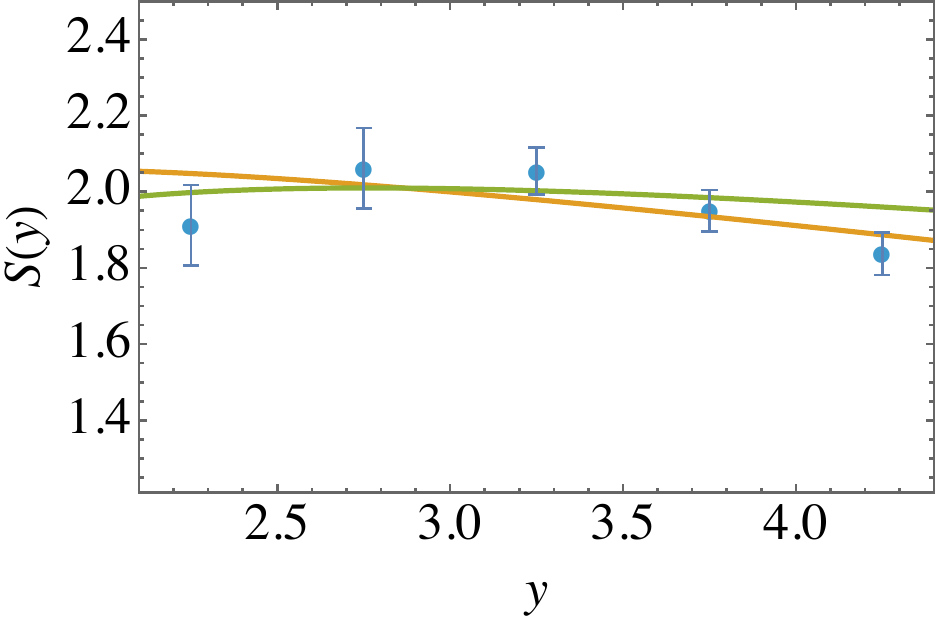}
  \end{minipage}~~
 \begin{minipage}[b]{0.45\textwidth}
    \centering
    \includegraphics[width=\linewidth]{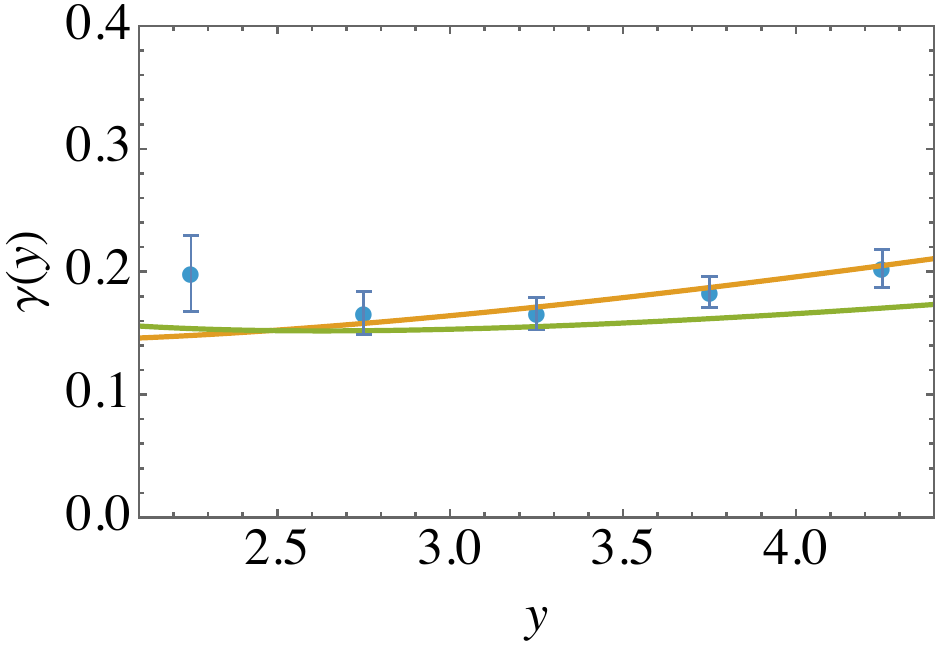}
  \end{minipage}\hfill
  \caption{Entropy (left panel) and purity (right panel)  obtained from the $\gamma$-cascade (solid lines)
  vs. data from Table~\ref{tab:nave_S}. Parameters of the $\gamma$-cascade
  were obtained
  by minimizing the $\chi^2$ for entropy (see text).
  Purity is therefore a prediction.}
  \label{fig:EPplot}
\end{figure}

In this section, we address  hadronic entropy and purity, which can be derived from probability distributions measured by the LHCb 
collaboration \cite{LHCb:2014wmv} and compare them with partonic cascades discussed in the previous sections. 
See also Ref.~\cite{Lokos:2025cbu} for a recent attempt to describe this data.
The LHCb kinematics at  $\sqrt{s}=7$~TeV is covering the rapidity range of $2 \le y\le 4.5$.
Therefore, the  final state partons are sensitive to paronic densities with very asymmetric longitudinal momentum fractions 
in colliding  protons, {\em i.e.} effectively one of the protons can be considered as a target while the other one as a projectile. 
The cross section for such process can be described within Hybrid Factorization version of Color Glass Condensate \cite{Dumitru:2005gt}. 
Here, however, we wish to apply the cascade models discussed above, to test various elements, {\em i.e.} recombination and transition to vacuum.
The LHCb measured probability
distributions in rapidity intervals $\Delta y=0.5$.
The corresponding entropy and purity can be computed in each interval (we chose
for $y$ the central value) from Eqs.~(\ref{eq:Sdef}) and (\ref{eq:purdef}).
The results are shown in Tab.~\ref{tab:nave_S} and in Fig.~(\ref{fig:EPplot}).

By inspecting the data \cite{LHCb:2014wmv}  one sees that the hadronic multiplicity is a non-monotonous function of rapidity. 
Therefore, as can be seen from Table~\ref{tab:nave_S}, the  entropy  follows this behavior. Hence,  it can not be described within the original 1 dimensional dipole model 
(\ref{eq:Equation0}),
as well as by the model accounting for merging of dipoles (\ref{eq:EquationSat1}). As we have shown in Sect.~\ref{sec:beta}
dipole mergers stabilize the increase of entropy, but cannot make it decrease with rapidity.
However, the transitions to vacuum or to states that escape the measurement,
which are modeled by the $\gamma$  terms (\ref{eq:eqsat2}), allow for a decrease of entropy.

\begin{table}[h]
    \centering
    \begin{tabular}{|c|c|c|}
    \hline
       $y$  & $S(y)$ & $\gamma(y)$  \\
       \hline
        2.25    &   $1.91\pm 0.11$ & $0.19\pm 0.03$\\
        2.75    &   $2.06\pm 0.11$ & $0.16\pm 0.017$\\
        3.25    &   $2.05\pm 0.06$ & $0.16\pm 0.013$\\
        3.75    &   $1.95\pm 0.06$ & $0.18\pm 0.012$\\
        4.25    &   $1.84\pm 0.06$ & $0.2\pm 0.015$ \\
        \hline
    \end{tabular}
    \caption{Average entropy and purity computed from the probability
    distributions measured by the LHCb Collaboration \cite{LHCb:2014wmv}.}
    \label{tab:nave_S}
\end{table}

In order to check whether the $\gamma$-cascade can reproduce general features of
the LHCb entropy and purity data given 
in Table~\ref{tab:nave_S}, we have performed two fits. In both cases we fitted entropy. In the first case
we  used all five LHCb points from Tab.~\ref{tab:nave_S} with the following result: $\alpha=2,~\beta=0.074$ and $\gamma=0.26$
with $\chi^2/{\rm n.d.f}\simeq 2$.
In the second fit we used four points  neglecting the point at largest rapidity.\footnote{During the fitting we have observed that
the parameter subspace ($\beta,\gamma$) has flat directions that may direct fits into unphysical regions.} 
The values of fitted parameters 
are $\alpha=1.2$, $\beta=0.02$, $\gamma=0.18$ with $\chi^2/{\rm n.d.f}\simeq 2$.
The values of the parameters indicate that the dominant processes are production and transitions to undetected states. The smallness 
of  merging is consistent with the theory requirement that such processes are less probable than the splittings, and the fact that $\gamma>\beta$
indicates that the system did not saturate. The results are shown in the left panel of Fig.~\ref{fig:EPplot} where the orange
line corresponds to the 5-point fit, whereas the green line to the fit with 4 points.

Having fixed the $\gamma$-cascade parameters we obtain {\em prediction} for the purity, which is shown in the right panel
of Fig.~\ref{fig:EPplot}. Interestingly, for the 5-point fit (shown as an orange line) the $\chi^2/{\rm n.d.f}\simeq 1.5$ is better than 
in the case of entropy. On the contrary, the value of $\chi^2$ for the 4-point fit (shown in green) is much worse: 6.3.

This is an interesting result: although the description of purity is not perfect once the entropy is described (and vice versa), apriori we did not expect the description to be so compatible with the data. Both measures  — entropy and purity — originate from the same multiplicity distribution, but there was no guarantee that a model reproducing one will automatically reproduce the other. This is most likely due to the fact that numerically
most important contributions to entropy and purity come
from different regions of $n$.



\section{Summary and conclusions}
\label{sec:conclusions}

Multiparticle production at high-energies a is a complex process that involves both perturbative and non-perturbative QCD physics.
It involves parton production  in a boosted projectile (or target), which is amenable to perturbative or semi-perturbative description
in terms of perturbative QCD, and non-perturbative hadronization processes, which are subject to phenomenological modeling.
Multiplicity distributions of hadrons and partons are closely related, but the exact relationship is subject to large theoretical 
uncertainties arising, for example, from the fact that QCD predicts in most cases gluon distributions, while experimentally 
only charged hadrons are observed. Nevertheless, the main features of both parton and hadron distributions should be in general
very similar. In contrast,  statistical measures of the multiparticle final states, such as entropy, are not affected by 
the hadronization processes. 

Implementation of the QCD cascades in the case of hadron-hadron scattering is a complicated task requiring involved analytical
computations and time consuming computer simulations \cite{Sjostrand:2006za,Bewick:2023tfi,CASCADE:2021bxe,Mueller:1996te,Domine:2018myf}. However, the general features of these processes can be understood 
in terms of much simpler, one-dimensional "toy models" that allow us to identify the fundamental physical 
phenomena responsible for parton production and recombination without the complications of four-dimensional QCD cascades.
In this paper we studied dipole cascade models that are motivated by the high energy limit of pQCD.
These cascades are formulated in 1 dimension and describe evolution of the dipoles - gluons with rapidity. 
While they have been studied already for some time, our careful numerical analyses have revealed several 
new features that we believe are interesting in their own right.

We have studied three types of cascades described by Eqs.(\ref{eq:Equation0}), 
(\ref{eq:EquationSat1}) 
and (\ref{eq:eqsat2})
called $\alpha$, $\beta$ and $\gamma-$cascades, respectively. The simplest $\alpha-$cascade (\ref{eq:Equation0})
can be solved analytically and leads to the untamed production of dipoles with geometric multiplicty distribution (\ref{eq:pneqPnG}).
Therefore, it is typically supplemented by the recombination $\beta-$terms (\ref{eq:EquationSat1}), which lead to the 
saturation of multiplicity. The solution of this previously known equation with splittings and recombinations can be characterized 
by the ratio of the couplings $r=\beta/\alpha$ corresponding to the splitting and recombination terms. We have identified
two regimes of the $\beta-$cascade, for large and small $r$, which we call "parallel" and "focal", respectively. In both regimes
the probability distribution of the cascade is very well described by the NBD, whose asymptotic properties such as multiplicity, dispersion
and the NBD $k$ parameter are determined solely by the value of $r$. However, only in the focal regime the scaling laws take
the expected very simple analytical form (\ref{eq:satscaling}).

This analysis shows that the $\beta-$cascade is not able to reproduce the energy dependence of the NBD $k$ parameter
which is decreasing with energy at the LHC \cite{Praszalowicz:2011zza}. It is also unable to describe the measurement of 
the multiplicity of charged hadrons  at forward rapidity at LHCb, which is not saturating but decreasing with rapidity.
 Therefore, we have introduced a new model that takes into account dipole-to-vacuum transitions, so called $\gamma-$cascade. The 
multiplicity of the  $\gamma-$cascade is indeed decreasing with rapidity. This is due to the fact, that $p_0$, which
is identically zero in the case of $\alpha$ and $\beta-$cascades, increases with $y$ removing  partons from the measured spectrum.
It turns out  that the probability distribution after removing $p_0$ can be also represented as a NBD with, however, $k$ parameter
which saturates for large $y$. Therefore the $\gamma-$cascade is not able to reproduce the energy dependence of the multiparicle
spectra at the LHC.

Next, we analyzed the properties of the cascades using tools motivated by quantum information theory, such as entropy, complexity, variance, 
and purity. These measures allow us to clearly demonstrate how the cascades differ and how the local dynamics was reflected 
in the properties of the solutions. 
In particular, we demonstrated that the $\beta$-cascade leads to the saturation of entropy, while the $\gamma$-cascade leads to a non-monotonic 
behavior of entropy: the entropy reaches a maximum and then drops.

Finally, we addressed the description of entropy data as inferred from multiplicity of charged hadrons measured by the LHCb
in the forward direction. Using the hadronic entropy as obtained from the multiplicity distributions we constrained parameters of 
the $\gamma$ cascade and used it to describe the purity data. 
This result confirms earlier conjecture by \cite{Kharzeev:2017qzs} and evidence by comparison of theory calculations to data \cite{Tu:2019ouv,Hentschinski:2022evidence,Hentschinski:2023maxent,Hentschinski:2024qcd_evo} that entropy and 
also purity are quantities that, although obtained for partonic degrees of freedom, can also characterize hadrons.
One can also reverse the logic and state that by calculating entropy and purity of data one can have direct information 
about partonic degrees of freedom.

\section*{Acknowledgements}

The authors acknowledge the hospitality and support of the Centro de Ciencias de Benasque Pedro Pascual, Spain, where part of this work was completed.
KK acknowledges valuable discussions with Martin Hentschinski and gratefully acknowledges the hospitality of the QCD group at École Polytechnique, Institut Polytechnique de Paris, where this work was completed. The work of KK was supported by NCN grant No. 2019/33/B/ST2/02588 and by a fellowship of 
the French Embassy in Warsaw.

\bibliographystyle{jhep} 
\bibliography{references}

\providecommand{\href}[2]{#2}\begingroup\raggedright\begin{thebibliography}{10}

\bibitem{Afik:2025ejh}
Y.~Afik et~al., \emph{{Quantum Information meets High-Energy Physics: Input to
  the update of the European Strategy for Particle Physics}},
  \href{https://arxiv.org/abs/2504.00086}{{\ttfamily 2504.00086}}.

\bibitem{DelGratta:2025qyp}
M.~Del~Gratta, F.~Fabbri, P.~Lamba, F.~Maltoni and D.~Pagani, \emph{{Quantum
  properties of $H\to VV^*$: precise predictions in the SM and sensitivity to
  new physics}},  \href{https://arxiv.org/abs/2504.03841}{{\ttfamily
  2504.03841}}.

\bibitem{STAR:2025njp}
{\scshape STAR} collaboration, \emph{{Probing QCD Confinement with Spin
  Entanglement}},  \href{https://arxiv.org/abs/2506.05499}{{\ttfamily
  2506.05499}}.

\bibitem{Maltoni:2024csn}
F.~Maltoni, C.~Severi, S.~Tentori and E.~Vryonidou, \emph{{Quantum tops at
  circular lepton colliders}},
  \href{http://dx.doi.org/10.1007/JHEP09(2024)001}{\emph{JHEP} {\bfseries 09}
  (2024) 001}, [\href{https://arxiv.org/abs/2404.08049}{{\ttfamily
  2404.08049}}].

\bibitem{Aoude:2023hxv}
R.~Aoude, E.~Madge, F.~Maltoni and L.~Mantani, \emph{{Probing new physics
  through entanglement in diboson production}},
  \href{http://dx.doi.org/10.1007/JHEP12(2023)017}{\emph{JHEP} {\bfseries 12}
  (2023) 017}, [\href{https://arxiv.org/abs/2307.09675}{{\ttfamily
  2307.09675}}].

\bibitem{Datta:2024hpn}
J.~Datta, A.~Deshpande, D.~E. Kharzeev, C.~J. Na{\"\i}m and Z.~Tu,
  \emph{{Entanglement as a Probe of Hadronization}},
  \href{http://dx.doi.org/10.1103/PhysRevLett.134.111902}{\emph{Phys. Rev.
  Lett.} {\bfseries 134} (2025) 111902},
  [\href{https://arxiv.org/abs/2410.22331}{{\ttfamily 2410.22331}}].

\bibitem{Florio:2025hoc}
A.~Florio, D.~Frenklakh, S.~Grieninger, D.~E. Kharzeev, A.~Palermo and S.~Shi,
  \emph{{Thermalization from quantum entanglement: jet simulations in the
  massive Schwinger model}},
  \href{https://arxiv.org/abs/2506.14983}{{\ttfamily 2506.14983}}.

\bibitem{Qi:2025onf}
W.~Qi, Z.~Guo and B.-W. Xiao, \emph{{Studying Maximal Entanglement and Bell
  Nonlocality at an Electron-Ion Collider}},
  \href{https://arxiv.org/abs/2506.12889}{{\ttfamily 2506.12889}}.

\bibitem{Hatta:2024lbw}
Y.~Hatta and J.~Montgomery, \emph{{Maximally entangled gluons for any x}},
  \href{http://dx.doi.org/10.1103/PhysRevD.111.014024}{\emph{Phys. Rev. D}
  {\bfseries 111} (2025) 014024},
  [\href{https://arxiv.org/abs/2410.16082}{{\ttfamily 2410.16082}}].

\bibitem{Altomonte:2024upf}
C.~Altomonte, A.~J. Barr, M.~Eckstein, P.~Horodecki and K.~Sakurai,
  \emph{{Prospects for quantum process tomography at high energies}},
  \href{https://arxiv.org/abs/2412.01892}{{\ttfamily 2412.01892}}.

\bibitem{Kharzeev:2017qzs}
D.~E. Kharzeev and E.~M. Levin, \emph{{Deep inelastic scattering as a probe of
  entanglement}},
  \href{http://dx.doi.org/10.1103/PhysRevD.95.114008}{\emph{Phys. Rev. D}
  {\bfseries 95} (2017) 114008},
  [\href{https://arxiv.org/abs/1702.03489}{{\ttfamily 1702.03489}}].

\bibitem{Mueller:1993rr}
A.~H. Mueller, \emph{{Soft gluons in the infinite momentum wave function and
  the BFKL pomeron}},
  \href{http://dx.doi.org/10.1016/0550-3213(94)90116-3}{\emph{Nucl. Phys. B}
  {\bfseries 415} (1994) 373--385}.

\bibitem{Mueller:1996te}
A.~H. Mueller and G.~P. Salam, \emph{{Large multiplicity fluctuations and
  saturation effects in onium collisions}},
  \href{http://dx.doi.org/10.1016/0550-3213(96)00336-7}{\emph{Nucl. Phys. B}
  {\bfseries 475} (1996) 293},
  [\href{https://arxiv.org/abs/hep-ph/9605302}{{\ttfamily hep-ph/9605302}}].

\bibitem{Liu:2022hto}
Y.~Liu, M.~A. Nowak and I.~Zahed, \emph{{Rapidity evolution of the entanglement
  entropy in quarkonium: Parton and string duality}},
  \href{http://dx.doi.org/10.1103/PhysRevD.105.114028}{\emph{Phys. Rev. D}
  {\bfseries 105} (2022) 114028},
  [\href{https://arxiv.org/abs/2203.00739}{{\ttfamily 2203.00739}}].

\bibitem{LevinLublinsky:2003Linear}
E.~Levin and M.~Lublinsky, \emph{A linear evolution for non-linear dynamics and
  correlations in realistic nuclei},
  \href{http://dx.doi.org/10.1016/j.nuclphysa.2003.10.020}{\emph{Nucl.\ Phys.\
  A} {\bfseries 730} (2004) 191--211},
  [\href{https://arxiv.org/abs/hep-ph/0308279}{{\ttfamily hep-ph/0308279}}].

\bibitem{Hentschinski:2022evidence}
M.~Hentschinski and K.~Kutak, \emph{Evidence for the maximally entangled low
  $x$ proton in deep inelastic scattering from h1 data},
  \href{http://dx.doi.org/10.1140/epjc/s10052-022-10056-y}{\emph{Eur.\ Phys.\
  J.\ C} {\bfseries 82} (2022) },
  [\href{https://arxiv.org/abs/2110.06156}{{\ttfamily 2110.06156}}].

\bibitem{Hentschinski:2022maxentDIS}
M.~Hentschinski, K.~Kutak and R.~Straka, \emph{Maximally entangled proton and
  charged hadron multiplicity in deep inelastic scattering},
  \href{http://dx.doi.org/10.1140/epjc/s10052-022-11122-1}{\emph{Eur.\ Phys.\
  J.\ C} {\bfseries 82} (2022) 1147}.

\bibitem{Hentschinski:2024qcd_evo}
M.~Hentschinski, D.~E. Kharzeev, K.~Kutak and Z.~Tu, \emph{Qcd evolution of
  entanglement entropy}, {\emph{Rep.\ Prog.\ Phys.} {\bfseries 87} (2024)
  120501}, [\href{https://arxiv.org/abs/2408.01259}{{\ttfamily 2408.01259}}].

\bibitem{Hentschinski:2023maxent}
M.~Hentschinski, D.~E. Kharzeev, K.~Kutak and Z.~Tu, \emph{Probing the onset of
  maximal entanglement inside the proton in diffractive deep inelastic
  scattering},
  \href{http://dx.doi.org/10.1103/PhysRevLett.131.241901}{\emph{Phys.\ Rev.\
  Lett.} {\bfseries 131} (2023) 241901}.

\bibitem{Tu:2019ouv}
Z.~Tu, D.~E. Kharzeev and T.~Ullrich, \emph{{Einstein-Podolsky-Rosen Paradox
  and Quantum Entanglement at Subnucleonic Scales}},
  \href{http://dx.doi.org/10.1103/PhysRevLett.124.062001}{\emph{Phys. Rev.
  Lett.} {\bfseries 124} (2020) 062001},
  [\href{https://arxiv.org/abs/1904.11974}{{\ttfamily 1904.11974}}].

\bibitem{Kutak:2011rb}
K.~Kutak, \emph{{Gluon saturation and entropy production in
  proton{\textendash}proton collisions}},
  \href{http://dx.doi.org/10.1016/j.physletb.2011.09.113}{\emph{Phys. Lett. B}
  {\bfseries 705} (2011) 217--221},
  [\href{https://arxiv.org/abs/1103.3654}{{\ttfamily 1103.3654}}].

\bibitem{Kutak:2023cwg}
K.~Kutak, \emph{{Entanglement entropy of proton and its relation to
  thermodynamics entropy}},  \href{https://arxiv.org/abs/2310.18510}{{\ttfamily
  2310.18510}}.

\bibitem{Gursoy:2023hge}
U.~G{\"u}rsoy, D.~E. Kharzeev and J.~F. Pedraza, \emph{{Universal rapidity
  scaling of entanglement entropy inside hadrons from conformal invariance}},
  \href{http://dx.doi.org/10.1103/PhysRevD.110.074008}{\emph{Phys. Rev. D}
  {\bfseries 110} (2024) 074008},
  [\href{https://arxiv.org/abs/2306.16145}{{\ttfamily 2306.16145}}].

\bibitem{Peschanski:2012cw}
R.~Peschanski, \emph{{Dynamical entropy of dense QCD states}},
  \href{http://dx.doi.org/10.1103/PhysRevD.87.034042}{\emph{Phys. Rev. D}
  {\bfseries 87} (2013) 034042},
  [\href{https://arxiv.org/abs/1211.6911}{{\ttfamily 1211.6911}}].

\bibitem{Stoffers:2012mn}
A.~Stoffers and I.~Zahed, \emph{Holographic entropy and real-time dynamics of
  quarkonium dissociation in qcd},
  \href{http://dx.doi.org/10.1103/PhysRevD.87.075023}{\emph{Phys. Rev. D}
  {\bfseries 87} (2013) 075023},
  [\href{https://arxiv.org/abs/1205.3223}{{\ttfamily 1205.3223}}].

\bibitem{Gotsman:2020bjc}
E.~Gotsman and E.~Levin, \emph{{High energy QCD: multiplicity distribution and
  entanglement entropy}},
  \href{http://dx.doi.org/10.1103/PhysRevD.102.074008}{\emph{Phys. Rev. D}
  {\bfseries 102} (2020) 074008},
  [\href{https://arxiv.org/abs/2006.11793}{{\ttfamily 2006.11793}}].

\bibitem{Kovner:2015hga}
A.~Kovner, M.~Lublinsky and V.~Skokov, \emph{Entanglement entropy, entropy
  production and time evolution in high energy qcd},
  \href{http://dx.doi.org/10.1103/PhysRevD.92.034045}{\emph{Phys. Rev. D}
  {\bfseries 92} (2015) 034045},
  [\href{https://arxiv.org/abs/1506.05394}{{\ttfamily 1506.05394}}].

\bibitem{Kovner:2018rbf}
A.~Kovner and M.~Lublinsky, \emph{Entanglement entropy and parton distributions
  in the color glass condensate framework},
  \href{http://dx.doi.org/10.1103/PhysRevD.99.074028}{\emph{Phys. Rev. D}
  {\bfseries 99} (2019) 074028},
  [\href{https://arxiv.org/abs/1811.08507}{{\ttfamily 1811.08507}}].

\bibitem{Peschanski:2016hgk}
R.~Peschanski and S.~Seki, \emph{{Entanglement Entropy of Scattering
  Particles}},
  \href{http://dx.doi.org/10.1016/j.physletb.2016.04.063}{\emph{Phys. Lett. B}
  {\bfseries 758} (2016) 89--92},
  [\href{https://arxiv.org/abs/1602.00720}{{\ttfamily 1602.00720}}].

\bibitem{Peschanski:2019yah}
R.~Peschanski and S.~Seki, \emph{{Evaluation of Entanglement Entropy in High
  Energy Elastic Scattering}},
  \href{http://dx.doi.org/10.1103/PhysRevD.100.076012}{\emph{Phys. Rev. D}
  {\bfseries 100} (2019) 076012},
  [\href{https://arxiv.org/abs/1906.09696}{{\ttfamily 1906.09696}}].

\bibitem{Dumitru:2023fih}
A.~Dumitru and E.~Kolbusz, \emph{{Quark pair angular correlations in the
  proton: Entropy versus entanglement negativity}},
  \href{http://dx.doi.org/10.1103/PhysRevD.108.034011}{\emph{Phys. Rev. D}
  {\bfseries 108} (2023) 034011},
  [\href{https://arxiv.org/abs/2303.07408}{{\ttfamily 2303.07408}}].

\bibitem{Dumitru:2023qee}
A.~Dumitru, A.~Kovner and V.~V. Skokov, \emph{{Entanglement entropy of the
  proton in coordinate space}},
  \href{http://dx.doi.org/10.1103/PhysRevD.108.014014}{\emph{Phys. Rev. D}
  {\bfseries 108} (2023) 014014},
  [\href{https://arxiv.org/abs/2304.08564}{{\ttfamily 2304.08564}}].

\bibitem{Ehlers:2022oal}
P.~J. Ehlers, \emph{{Entanglement between valence and sea quarks in hadrons of
  1+1 dimensional QCD}},
  \href{http://dx.doi.org/10.1016/j.aop.2023.169290}{\emph{Annals Phys.}
  {\bfseries 452} (2023) 169290},
  [\href{https://arxiv.org/abs/2209.09867}{{\ttfamily 2209.09867}}].

\bibitem{Dumitru:2025bib}
A.~Dumitru and E.~Kolbusz, \emph{{Quantum entanglement correlations in double
  quark PDFs}}, \href{http://dx.doi.org/10.1103/ngdx-pc85}{\emph{Phys. Rev. D}
  {\bfseries 111} (2025) 114033},
  [\href{https://arxiv.org/abs/2501.12312}{{\ttfamily 2501.12312}}].

\bibitem{Berges:2017hne}
J.~Berges, S.~Floerchinger and R.~Venugopalan, \emph{{Dynamics of entanglement
  in expanding quantum fields}},
  \href{http://dx.doi.org/10.1007/JHEP04(2018)145}{\emph{JHEP} {\bfseries 04}
  (2018) 145}, [\href{https://arxiv.org/abs/1712.09362}{{\ttfamily
  1712.09362}}].

\bibitem{Dvali:2021ooc}
G.~Dvali and R.~Venugopalan, \emph{{Classicalization and unitarization of wee
  partons in QCD and gravity: The CGC-black hole correspondence}},
  \href{http://dx.doi.org/10.1103/PhysRevD.105.056026}{\emph{Phys. Rev. D}
  {\bfseries 105} (2022) 056026},
  [\href{https://arxiv.org/abs/2106.11989}{{\ttfamily 2106.11989}}].

\bibitem{Chachamis:2023omp}
G.~Chachamis, M.~Hentschinski and A.~Sabio~Vera, \emph{{Von Neumann entropy and
  Lindblad decoherence in the high-energy limit of strong interactions}},
  \href{http://dx.doi.org/10.1103/PhysRevD.109.054015}{\emph{Phys. Rev. D}
  {\bfseries 109} (2024) 054015},
  [\href{https://arxiv.org/abs/2312.16743}{{\ttfamily 2312.16743}}].

\bibitem{Rybczynski:2025ccy}
M.~Rybczy{\'n}ski and Z.~W{\l}odarczyk, \emph{{Imprints of QCD cascades in
  hadron multiplicity distributions}},
  \href{http://dx.doi.org/10.1103/9mm7-rnxx}{\emph{Phys. Rev. D} {\bfseries
  111} (2025) 094045}, [\href{https://arxiv.org/abs/2504.11152}{{\ttfamily
  2504.11152}}].

\bibitem{Ramos:2025tge}
G.~S. Ramos, L.~S. Moriggi and M.~V.~T. Machado, \emph{{Investigating QCD
  dynamical entropy in high-energy nuclear collisions}},
  \href{http://dx.doi.org/10.1016/j.physletb.2025.139737}{\emph{Phys. Lett. B}
  {\bfseries 868} (2025) 139737},
  [\href{https://arxiv.org/abs/2507.09349}{{\ttfamily 2507.09349}}].

\bibitem{Cheskis:2025mrs}
D.~Cheskis and A.~Prygarin, \emph{{Pomeron Evolution and Squeezed States in
  Quantum Optics}},  \href{https://arxiv.org/abs/2505.02684}{{\ttfamily
  2505.02684}}.

\bibitem{Ouchen:2025tta}
M.~Ouchen and A.~Prygarin, \emph{{Pomeron evolution, entanglement entropy and
  Abramovskii-Gribov-Kancheli cutting rules}},
  \href{https://arxiv.org/abs/2508.12102}{{\ttfamily 2508.12102}}.

\bibitem{Iancu:2004iy}
E.~Iancu and D.~N. Triantafyllopoulos, \emph{{A Langevin equation for high
  energy evolution with pomeron loops}},
  \href{http://dx.doi.org/10.1016/j.nuclphysa.2005.03.124}{\emph{Nucl. Phys. A}
  {\bfseries 756} (2005) 419--467},
  [\href{https://arxiv.org/abs/hep-ph/0411405}{{\ttfamily hep-ph/0411405}}].

\bibitem{Iancu:2005dx}
E.~Iancu, G.~Soyez and D.~N. Triantafyllopoulos, \emph{{On the probabilistic
  interpretation of the evolution equations with Pomeron loops in QCD}},
  \href{http://dx.doi.org/10.1016/j.nuclphysa.2006.01.013}{\emph{Nucl. Phys. A}
  {\bfseries 768} (2006) 194--221},
  [\href{https://arxiv.org/abs/hep-ph/0510094}{{\ttfamily hep-ph/0510094}}].

\bibitem{Shoshi:2005pf}
A.~I. Shoshi and B.-W. Xiao, \emph{{Pomeron loops in zero transverse
  dimensions}}, \href{http://dx.doi.org/10.1103/PhysRevD.73.094014}{\emph{Phys.
  Rev. D} {\bfseries 73} (2006) 094014},
  [\href{https://arxiv.org/abs/hep-ph/0512206}{{\ttfamily hep-ph/0512206}}].

\bibitem{Kovner:2005aq}
A.~Kovner and M.~Lublinsky, \emph{{More remarks on high energy evolution}},
  \href{http://dx.doi.org/10.1016/j.nuclphysa.2005.12.010}{\emph{Nucl. Phys. A}
  {\bfseries 767} (2006) 171--188},
  [\href{https://arxiv.org/abs/hep-ph/0510047}{{\ttfamily hep-ph/0510047}}].

\bibitem{Kozlov:2006zj}
M.~Kozlov and E.~Levin, \emph{{Solution for the BFKL Pomeron Calculus in zero
  transverse dimensions}},
  \href{http://dx.doi.org/10.1016/j.nuclphysa.2006.08.011}{\emph{Nucl. Phys. A}
  {\bfseries 779} (2006) 142--176},
  [\href{https://arxiv.org/abs/hep-ph/0604039}{{\ttfamily hep-ph/0604039}}].

\bibitem{Bondarenko:2006rh}
S.~Bondarenko, L.~Motyka, A.~H. Mueller, A.~I. Shoshi and B.~W. Xiao, \emph{{On
  the equivalence of Reggeon field theory in zero transverse dimensions and
  reaction-diffusion processes}},
  \href{http://dx.doi.org/10.1140/epjc/s10052-007-0218-6}{\emph{Eur. Phys. J.
  C} {\bfseries 50} (2007) 593--601},
  [\href{https://arxiv.org/abs/hep-ph/0609213}{{\ttfamily hep-ph/0609213}}].

\bibitem{Kovner:2024tin}
A.~Kovner, E.~Levin and M.~Lublinsky, \emph{{High energy scattering in the
  Unitary Toy Model}},
  \href{http://dx.doi.org/10.1007/JHEP10(2024)127}{\emph{JHEP} {\bfseries 10}
  (2024) 127}, [\href{https://arxiv.org/abs/2406.12691}{{\ttfamily
  2406.12691}}].

\bibitem{Braun:2024dfr}
M.~A. Braun, \emph{{Entropy in Toy Regge models}},
  \href{https://arxiv.org/abs/2409.01620}{{\ttfamily 2409.01620}}.

\bibitem{Kou:2022dkw}
W.~Kou, X.~Wang and X.~Chen, \emph{{Page entropy of a proton system in deep
  inelastic scattering at small x}},
  \href{http://dx.doi.org/10.1103/PhysRevD.106.096027}{\emph{Phys. Rev. D}
  {\bfseries 106} (2022) 096027},
  [\href{https://arxiv.org/abs/2208.07521}{{\ttfamily 2208.07521}}].

\bibitem{Giovannini:1985mz}
A.~Giovannini and L.~Van~Hove, \emph{{Negative Binomial Multiplicity
  Distributions in High-Energy Hadron Collisions}},
  \href{http://dx.doi.org/10.1007/BF01557602}{\emph{Z. Phys. C} {\bfseries 30}
  (1986) 391}.

\bibitem{Praszalowicz:2011zza}
M.~Praszalowicz, \emph{{Negative Binomial Distribution and the multiplicity
  moments at the LHC}},
  \href{http://dx.doi.org/10.1016/j.physletb.2011.09.101}{\emph{Phys. Lett. B}
  {\bfseries 704} (2011) 566--569},
  [\href{https://arxiv.org/abs/1101.6012}{{\ttfamily 1101.6012}}].

\bibitem{Gelis:2009wh}
F.~Gelis, T.~Lappi and L.~McLerran, \emph{{Glittering Glasmas}},
  \href{http://dx.doi.org/10.1016/j.nuclphysa.2009.07.004}{\emph{Nucl. Phys. A}
  {\bfseries 828} (2009) 149--160},
  [\href{https://arxiv.org/abs/0905.3234}{{\ttfamily 0905.3234}}].

\bibitem{DiasdeDeus:2010ggs}
J.~Dias~de Deus and C.~Pajares, \emph{{String Percolation and the Glasma}},
  \href{http://dx.doi.org/10.1016/j.physletb.2010.11.017}{\emph{Phys. Lett. B}
  {\bfseries 695} (2011) 211--213},
  [\href{https://arxiv.org/abs/1011.1099}{{\ttfamily 1011.1099}}].

\bibitem{Caputa:2024xkp}
P.~Caputa and K.~Kutak, \emph{{Krylov complexity and gluon cascades in the high
  energy limit}},
  \href{http://dx.doi.org/10.1103/PhysRevD.110.085011}{\emph{Phys. Rev. D}
  {\bfseries 110} (2024) 085011},
  [\href{https://arxiv.org/abs/2404.07657}{{\ttfamily 2404.07657}}].

\bibitem{Kovchegov:2012mbw}
Y.~V. Kovchegov and E.~Levin, \emph{{Quantum Chromodynamics at High Energy}},
  vol.~33.
\newblock Oxford University Press, 2013,
  \href{http://dx.doi.org/10.1017/9781009291446}{10.1017/9781009291446}.

\bibitem{Balitsky:1978ic}
I.~I. Balitsky and L.~N. Lipatov, \emph{{The Pomeranchuk Singularity in Quantum
  Chromodynamics}}, {\emph{Sov. J. Nucl. Phys.} {\bfseries 28} (1978)
  822--829}.

\bibitem{Kuraev:1977fs}
E.~A. Kuraev, L.~N. Lipatov and V.~S. Fadin, \emph{{The Pomeranchuk Singularity
  in Nonabelian Gauge Theories}}, {\emph{Sov. Phys. JETP} {\bfseries 45} (1977)
  199--204}.

\bibitem{Balitsky:1995ub}
I.~Balitsky, \emph{{Operator expansion for high-energy scattering}},
  \href{http://dx.doi.org/10.1016/0550-3213(95)00638-9}{\emph{Nucl. Phys. B}
  {\bfseries 463} (1996) 99--160},
  [\href{https://arxiv.org/abs/hep-ph/9509348}{{\ttfamily hep-ph/9509348}}].

\bibitem{Kovchegov:1999yj}
Y.~V. Kovchegov, \emph{{Small x F(2) structure function of a nucleus including
  multiple pomeron exchanges}},
  \href{http://dx.doi.org/10.1103/PhysRevD.60.034008}{\emph{Phys. Rev. D}
  {\bfseries 60} (1999) 034008},
  [\href{https://arxiv.org/abs/hep-ph/9901281}{{\ttfamily hep-ph/9901281}}].

\bibitem{Mueller1995UnitarityBFKL}
A.~H. Mueller, \emph{Unitarity and the bfkl pomeron},
  \href{http://dx.doi.org/10.1016/0550-3213(94)00480-3}{\emph{Nuclear Physics
  B} {\bfseries 437} (1995) 107--126},
  [\href{https://arxiv.org/abs/hep-ph/9408245}{{\ttfamily hep-ph/9408245}}].

\bibitem{Szwed:1987vj}
R.~Szwed, G.~Wrochna and A.~K. Wroblewski, \emph{{Mystery of the Negative
  Binomial Distribution}}, {\emph{Acta Phys. Polon. B} {\bfseries 19} (1988)
  763}.

\bibitem{Hagiwara:2017uaz}
Y.~Hagiwara, Y.~Hatta, B.-W. Xiao and F.~Yuan, \emph{{Classical and quantum
  entropy of parton distributions}},
  \href{http://dx.doi.org/10.1103/PhysRevD.97.094029}{\emph{Phys. Rev. D}
  {\bfseries 97} (2018) 094029},
  [\href{https://arxiv.org/abs/1801.00087}{{\ttfamily 1801.00087}}].

\bibitem{Blaizot:2006wp}
J.~P. Blaizot, E.~Iancu and D.~N. Triantafyllopoulos, \emph{{A Zero-dimensional
  model for high-energy scattering in QCD}},
  \href{http://dx.doi.org/10.1016/j.nuclphysa.2006.11.127}{\emph{Nucl. Phys. A}
  {\bfseries 784} (2007) 227--258},
  [\href{https://arxiv.org/abs/hep-ph/0606253}{{\ttfamily hep-ph/0606253}}].

\bibitem{Baiguera:2025dkc}
S.~Baiguera, V.~Balasubramanian, P.~Caputa, S.~Chapman, J.~Haferkamp, M.~P.
  Heller et~al., \emph{{Quantum complexity in gravity, quantum field theory,
  and quantum information science}},
  \href{https://arxiv.org/abs/2503.10753}{{\ttfamily 2503.10753}}.

\bibitem{LHCb:2014wmv}
{\scshape LHCb} collaboration, R.~Aaij et~al., \emph{{Measurement of charged
  particle multiplicities and densities in $pp$ collisions at $\sqrt{s}=7\;$TeV
  in the forward region}},
  \href{http://dx.doi.org/10.1140/epjc/s10052-014-2888-1}{\emph{Eur. Phys. J.
  C} {\bfseries 74} (2014) 2888},
  [\href{https://arxiv.org/abs/1402.4430}{{\ttfamily 1402.4430}}].

\bibitem{Lokos:2025cbu}
S.~L{\"o}k{\"o}s, \emph{{Charged-particle Multiplicity Distributions Derived
  from the Principle of Maximal Entropy}},
  \href{http://dx.doi.org/10.5506/APhysPolBSupp.18.5-A19}{\emph{Acta Phys.
  Polon. Supp.} {\bfseries 18} (2025) 5--A19},
  [\href{https://arxiv.org/abs/2505.23491}{{\ttfamily 2505.23491}}].

\bibitem{Dumitru:2005gt}
A.~Dumitru, A.~Hayashigaki and J.~Jalilian-Marian, \emph{{The Color glass
  condensate and hadron production in the forward region}},
  \href{http://dx.doi.org/10.1016/j.nuclphysa.2005.11.014}{\emph{Nucl. Phys. A}
  {\bfseries 765} (2006) 464--482},
  [\href{https://arxiv.org/abs/hep-ph/0506308}{{\ttfamily hep-ph/0506308}}].

\bibitem{Sjostrand:2006za}
T.~Sjostrand, S.~Mrenna and P.~Z. Skands, \emph{{PYTHIA 6.4 Physics and
  Manual}}, \href{http://dx.doi.org/10.1088/1126-6708/2006/05/026}{\emph{JHEP}
  {\bfseries 05} (2006) 026},
  [\href{https://arxiv.org/abs/hep-ph/0603175}{{\ttfamily hep-ph/0603175}}].

\bibitem{Bewick:2023tfi}
G.~Bewick et~al., \emph{{Herwig 7.3 release note}},
  \href{http://dx.doi.org/10.1140/epjc/s10052-024-13211-9}{\emph{Eur. Phys. J.
  C} {\bfseries 84} (2024) 1053},
  [\href{https://arxiv.org/abs/2312.05175}{{\ttfamily 2312.05175}}].

\bibitem{CASCADE:2021bxe}
{\scshape CASCADE} collaboration, S.~Baranov et~al., \emph{{CASCADE3 A Monte
  Carlo event generator based on TMDs}},
  \href{http://dx.doi.org/10.1140/epjc/s10052-021-09203-8}{\emph{Eur. Phys. J.
  C} {\bfseries 81} (2021) 425},
  [\href{https://arxiv.org/abs/2101.10221}{{\ttfamily 2101.10221}}].

\bibitem{Domine:2018myf}
L.~Domin{\'e}, G.~Giacalone, C.~Lorc{\'e}, S.~Munier and S.~Pekar, \emph{{Gluon
  density fluctuations in dilute hadrons}},
  \href{http://dx.doi.org/10.1103/PhysRevD.98.114032}{\emph{Phys. Rev. D}
  {\bfseries 98} (2018) 114032},
  [\href{https://arxiv.org/abs/1810.05049}{{\ttfamily 1810.05049}}].

\end{thebibliography}\endgroup

\end{document}